\documentclass[aps,prb,amsmath,amssymb,twocolumn,reprint,showpacs,longbibliography]{revtex4-2}
\pdfoutput=1
\bibpunct{[}{]}{,}{n}{}{}

\usepackage[utf8]{inputenc}
\usepackage{bm,latexsym,mathrsfs,enumerate}
\usepackage[dvipsnames]{xcolor}
\usepackage{mathtools}
\usepackage{makecell}
\usepackage{multirow}
\usepackage[breaklinks=true,unicode=true,urlcolor = blue,colorlinks = true,citecolor = blue,linkcolor = blue]{hyperref}
\usepackage{graphicx}
\usepackage{ragged2e}
\usepackage[export]{adjustbox}

\renewcommand{\vec}[1]{\bm{#1}}
\usepackage[final]{pdfpages}
\usepackage{tikz}
\makeatletter
\AtBeginDocument{\let\LS@rot\@undefined}
\makeatother

\begin{document}

\title{Ubiquity of the spin-orbit induced magnon nonreciprocity in ultrathin ferromagnets}%

\author{Albrecht von Faber} 
\author{Christopher Hins}
\author{Khalil Zakeri}\email{khalil.zakeri@partner.kit.edu}

\affiliation{Heisenberg Spin-dynamics Group, Physikalisches Institut,
Karlsruhe Institute of Technology, Wolfgang-Gaede-Strasse 1, D-76131 Karlsruhe, Germany}%

\date{\today}%

\begin{abstract}
  The propagation of magnons along a symmetry path may depend on the direction of propagation, similar to many other quasiparticles in nature. This phenomenon is commonly referred to as nonreciprocity. In addition to the fact that it is of great interest to understand the fundamental physical mechanism leading to this nonreciprocal propagation, the phenomenon of magnon nonreciprocity may be used to design magnon-based logic devices. Recently, it has been demonstrated that a significantly large spin-orbit coupling can lead to giant nonreciprocity of exchange-dominated terahertz magnons, when they are excited by means of spin-polarized electrons [\href{https://doi.org/10.1103/PhysRevLett.132.126702}{Phys.~Rev.~Lett.~\textbf{132},~126702~(2024)}].  Here, by providing experimental results on two additional systems we demonstrate the generality of the observed phenomenon. Comparing the results of a Co/Ni bilayer on Ir(001) to those of a Co double layer on Ir(001) and W(110) we unravel the impact of the interfacial electronic hybridization on the observed phenomenon and provide further insights into the microscopic mechanism leading to this nonreciprocal magnon excitation. It was observed that the interfacial electronic hybridization is of some importance but is not crucial for the magnon nonreciprocity. This is an important observation since the electronic hybridization is known to be a key aspect in the determination of the magnetic properties at the interface. On the other hand the choice of the incident energy of the incoming electron beam is decisive for the observation of the effect. Our results indicate that depending on the energy of the incident electron beam and the scattering geometry the magnon nonreciprocity can be tuned and even be inverted for some ranges of the magnon momentum. The results may inspire new ideas for the generation of nonreciprocal ultrafast terahertz magnons in ultrathin films of $3d$ magnetic metals as well as in two-dimensional van der Waals magnets, interfaced with solids possessing a large spin-orbit coupling, for applications in the field of magnonics.  
\end{abstract}

\maketitle


\section{Introduction}

Usually when a quasiparticle is propagating along two opposite but equivalent directions it is expected to exhibit identical properties, e.g., amplitude, lifetime, propagation speed, etc. \cite{Feng2011,Ramezani2018}. However, under some circumstances, e.g., in the presence of a chiral interaction, these properties may depend on the direction of propagation. This phenomenon is usually referred to as nonreciprocity and has been expected to be observed for magnons --the quanta of spin waves in magnetic materials-- when they are propagating along two opposite directions of a symmetry path. The effect is of great importance from both fundamental as well as application points of view. From a fundamental point of view, a nonreciprocal propagation of a quasiparticle is associated with the presence of an interaction, which possibly breaks the time-reversal symmetry. It is, therefore, interesting to unravel the nature of the nonreciprocity, in order to address the presence of such a time-reversal symmetry breaking interaction. From an application point of view, nonreciprocal propagation of magnons is of great importance to realize novel magnon-based logic devices useful for applications in the field of magnonics and spintronics, as has already been demonstrated in the case of magnons with gigahertz frequencies \cite{Damon1961,Gruenberg1985,Seki2016,Mruczkiewicz2017,Ishibashi2020,Wang2020, Dobrovolskiy2022,Gladii2023}.

From the band magnetism point of view magnons in itinerant electron systems can be described by the coherent superposition of all possible  electron-hole pairs for which the electron is of minority spin character and the hole is of majority character, formed across the Fermi-level. The wavefunction of such an excitation can be written as a linear superposition of the  wavefunctions of all possible correlated electron-hole pairs, which satisfy this condition (the fact that the total angular momentum of a magnon is $1\hbar$), as well as the energy and wavevector conservation rules \cite{Mills2007,Zakeri2014}. It is straightforward to imagine that if the degeneracy of the bands is broken solely due to the exchange splitting, one would not expect a difference in the amplitude of the magnons' wavefunction, in particular in the case of exchange-dominated magnons, with the opposite wavevector (or the two opposite directions of the magnetization,  if the sample is  ferromagnetic). The situation changes if one adds the relativistic spin-orbit coupling (SOC) into the formalism. In this case in addition to the exchange splitting one needs to consider the spin-orbit induced band splitting. Hence, one would expect a difference in the amplitude of the magnons' wavefunction with the opposite signs of  the wavevector (or for the two opposite directions of the magnetization). This is a direct consequence of the fact that the electrons (and holes) belong to spin-orbit spin-mixed bands now. In an extreme case scenario e.g., in the case of topological insulators, it has been predicted that due to the spin-momentum locking of the electronic states---a unique intrinsic feature of strongly spin-orbit coupled states in solids--- even the collective charge excitations shall carry a spin character \cite{Raghu2010,Zakeri2021b}.

From a broader perspective, one may consider a magnon as a bosonic quasiparticle which possesses three characteristics, i.~e., (i) an eigenfrequency,  (ii) a lifetime and (iii) an amplitude. In principle, SOC can lead to a magnon nonreciprocity in different ways.

First, SOC as a chiral interaction can break the energy  degeneracy of the magnons having the same wavevector but propagate along opposite (but crystallographically equivalent) directions and, thereby, can lead to nonreciprocal magnons. However, this interaction does not directly act on magnons as bosons. We have demonstrated earlier that  SOC leads to the presence of the antisymmetric Dzyaloshinskii--Moriya interaction (DMI), which can then influence the magnon eigenfrequency \cite{Zakeri2010}. In a simplified picture SOC appears in the spin Hamiltonian in the form of DMI \cite{Zakeri2010,Udvardi2009,Costa2010a,Costa2020}.

Second, SOC can influence the magnon lifetime by modifying the Landau damping of the magnons having opposite propagation directions. In metallic magnets the lifetime of high-energy exchange-dominated magnons is governed by their dissipation into single-particle electron-hole pairs (also called Stoner pairs), known as Landau damping \cite{Buczek2011,Qin2015,Zakeri2021a,Paischer2024}. Since in ultrathin ferromagnetic films grown on  heavy-element metallic substrates SOC inside the ferromagnetic film can be large, the dissipation  takes place within the spin-orbit spin-mixed electronic bands. It has been shown that this can lead to an asymmetric magnon lifetime \cite{Zakeri2012}. SOC leads to an additional decay channel, which depends on the propagation direction of the magnons \cite{Zakeri2012,Costa2010a}.

Third, SOC should, in principle, influence also the amplitude of the excited magnons via the excitation process, depending on the sign of their wavevector. This phenomenon has just been demonstrated very recently \cite{Zakeri2024}. We have shown that if the electrons are used as the excitation and probe particles, the phenomenon can be easily observed. The chiral SOC  leads to a nonreciprocal excitation cross-section and, hence, nonreciprocal magnon amplitudes. This is due to the fact that SOC is a chiral interaction and in ultrathin ferromagnets the time-reversal $\mathcal{T}$ and inversion symmetry is broken.
It is important to notice that such a nonreciprocity is caused by the fact that the excitation cross-section of magnons is not the same for the two opposite propagation directions (or the two opposite directions of magnetization). The effect is intimately related to the exchange process as the microscopic physical mechanism leading to the excitation of these exchange-dominated magnons.
Hence, this kind of nonreciprocity is expected to be observed in all excitation schemes, which involve the exchange mechanism as the underlying mechanism for magnon excitation.

It is worth emphasizing that the magnon nonreciprocity is expected to be observed for the exchange-dominated terahertz (THz) magnons. Owing to the ultrafast propagation of these type of magnons, they would allow for information processing on ultrashort time scales \cite{Zakeri2020}. Hence, such magnons are of great interest to the field of ultrafast magnonics and spintronics \cite{Zakeri2018}.

It has been observed, several decades ago, that when a longitudinally spin-polarized beam of electrons with the spin polarization vector $\vec{P}$ parallel and antiparallel to the scattering's plane normal vector $\vec{\hat{n}}$ is scattered from the surface of a nonmagnetic crystal with a large SOC, the intensity profile of the scattered electron beam can depend on the direction of $\vec{P}$. This fact leads to a nonzero  spin asymmetry, known as spin-orbit asymmetry (SOA) \cite{Wang1979,Kirschner1985,Feder1986} (see also Ref.~\cite{Zakeri2022} and references therein).
Likewise, in the case of ferromagnetic surfaces the exchange scattering process can also lead to a spin asymmetry, when $\vec{P}$ is parallel and antiparallel to the direction of the sample magnetization $\vec{M}$. The asymmetry observed in this case is a consequence of the  exchange scattering mechanism and is  known as exchange asymmetry (ExA)  \cite{Elmers2007}.
While SOC in bulk $3d$-ferromagnetic metals is rather weak, the observed spin asymmetry of the spin-polarized electrons scattered from such surfaces is predominantly of exchange type. However, when an ultrathin film of a ferromagnetic $3d$ metal is grown on a substrate with a strong SOC one would expect that both SOA and ExA are present  in the scattering event. Hence, one would expect to observe a competition between these two spin-dependent scattering mechanisms. On the other hand during the scattering event magnons may be excited, when all the necessary conditions are fulfilled (conservation law of energy, momentum and angular momentum). Since the excitation process of magnons is also based on the exchange scattering and is mediated by the fundamental electrostatic Coulomb interaction, one would expect that a competition between spin-orbit and exchange scattering processes may lead to a nonreciprocal excitation of magnons, when they are excited by means of spin-polarized electrons. 

In order to unravel the microscopic origin of the spin-orbit induced magnon nonreciprocity we investigated the impact  of two important factors on the magnon nonreciprocity: (i) the interfacial electronic hybridization, and (ii) the atomic structure and the symmetry of the magnetic overlayer as well as the choice of the substrate. 

In ferromagnetic films grown on heavy-element substrates, which are the main focus of the present work, it is generally accepted that  SOC is mainly an interfacial effect and originates from the underlying heavy-element metallic substrates. In this respect, it is important to know what is the role of the interfacial electronic hybridization in the observed effect. Note that the role of the interfacial electronic hybridization is known to be extremely important in the determination of  the magnetic properties of such ultrathin films. For instance, the interfacial magnetic moments and the Heisenberg exchange parameters depend strongly on the degree of the interfacial electronic hybridization \cite{Zakeri2017}. In connection to this point we have recently investigated the role of the interfacial electronic hybridization on the strength and chirality of DMI \cite{Zakeri2024b}. Comparing the experimental results to those of first principles calculations it was observed that such a hybridization is crucial for DMI (both its strength and chirality) \cite{Zakeri2024b}.

Although the spin-orbit induced magnon nonreciprocity is expected to be observed by other excitation schemes, it should be best investigated by means of spin-polarized electron scattering experiments, e.g., spin-polarized high-resolution electron energy-loss spectroscopy (SPHREELS).  While the original experiments were performed on a Co double layer on Ir(001), in order to investigate the impact of the interfacial electronic hybridization on the magnon nonreciprocity, we replaced the interface Co atomic layer with a Ni layer. We note that both Co double layer and Co/Ni bilayer \footnote{The term double layer refers to the case in which the sample is composed of two atomic layers of the same material. Likewise, the term bilayer refers to the case in which the sample is composed of two atomic layers made of two different materials.} possess nearly the same geometrical structure (crystallographic structure and the interatomic distances) \cite{Heinz2009,Zakeri2021,Zakeri2023b}.  Since most of the $3d$ states of Ni are occupied and are located below the Fermi level, it is expected that these states hybridize rather weakly with that of the $5d$ states of the Ir substrate \cite{Zakeri2024a,Zakeri2024b}. This would provide a unique possibility to  examine the influence of the interfacial electronic hybridization on the observed magnon nonreciprocity. On the other hand considering the available $5d$ states of the substrate, the tungsten exhibits a lower number of occupied $5d$ states compared to iridium. Hence, it would be interesting to replace the Ir(001) substrate with W(110) and see how it changes the magnon nonreciprocity. Unlike Ir, which possesses a face-centered cubic structure, the crystal structure of W is body-centered cubic. Since in the case of Co double layer on W(110) the atomic arrangement is also different, one would, therefore, expect to observe effects associated with the change in the atomic arrangement as well. In order to address the above mentioned points we investigated the details of the spin-dependent electron scattering from two systems, i.e., Co/Ni/Ir(001) and 2Co/W(110) and compare the results to those of 2Co/Ir(001), investigated previously \cite{Zakeri2024}.
We will show that, although the interfacial electronic hybridization is crucial for the determination the magnetic properties of the magnetic layers grown on a nonmagnetic substrate, they are not decisive for the observed magnon nonreciprocity. In contrary, the magnon nonreciprocity depends strongly on the incident electron energy and the scattering geometry. Moreover, for some ranges of magnon wavevectors the magnon nonreciprocity can even be inverted. 

The paper is organized as following. In Sec.~\ref{Sec:Experiment} we provide the experimental details. The substrate preparation, film growth and the SPHREELS experiments are explained. The results of the two systems, namely Co/Ni/Ir(001) and 2Co/W(110) are presented  in Secs.~\ref{Sec:Results_CoNi} and \ref{Sec:Results_2CoW}, respectively, and are discussed in detail. A concluding remark with the prospective of the observation of the effect in other layered magnetic structures is provides in Sec.~\ref{Sec:Conlusion}.

\section{Experimental details}\label{Sec:Experiment}

\subsection{Substrate preparation}
In this study we used Ir(001) and W(110) as substrates. The choice of Ir(001) is based on the fact that one can grow well-ordered ultrathin films of $3d$ ferromagnetic metals on this substrate. In particular considering Co and Ni as overlayers the growth mode, lattice structure and the interatomic distances are very similar \cite{Heinz2009,Tian2009,Qin2013}. They both grow in a face-centered tetragonal structure \cite{Heinz2009,Tian2009,Qin2013}. This is particularly important to address the effects associated with the changes in the chemical properties of the interface layer without changing its geometrical structures.
The choice of W(110) is based on the fact that it also exhibits a rather large SOC. Moreover, one can grow high-quality ultrathin films of $3d$ ferromagnets on this surface \cite{Pratzer2003,Etzkorn2005,Prokop2009,Zakeri2011,Tsurkan2020}.

Prior to the deposition of synthetic  layered structures, the surface of Ir(001) and W(110) was cleaned by cycles of flashing in oxygen atmosphere followed by flashing under ultra-high vacuum conditions.

In the case of the Ir(001) substrate the crystal was quickly heated up (flashed) multiple times to 1700~K under an oxygen partial pressure of $P_{O_2}$=$2\times10^{-7}$~mbar in order to remove the carbon contamination. The sample was then flashed at 2200~K to remove the oxygen. This leads to an atomically clean and well-ordered Ir(001)--$1\times5$ surface, as confirmed by low-energy electron diffraction patterns \cite{Qin2013,Chuang2014,Zakeri2021}.

In the case of  W(110) a similar procedure was used. The initial cleaning procedure was performed at a lower temperature (1200 K). The details of the cleaning procedure may be found in Ref.~\cite{Zakeri2010a}). The low-energy electron diffraction patterns recorded after this procedure indicated a contamination-free W(110)--$1\times1$ surface \cite{Etzkorn2005,Prokop2009,Zakeri2010a,Zakeri2011}. 

\begin{figure}[t!]
	\centering
	\includegraphics[width=0.95\columnwidth]{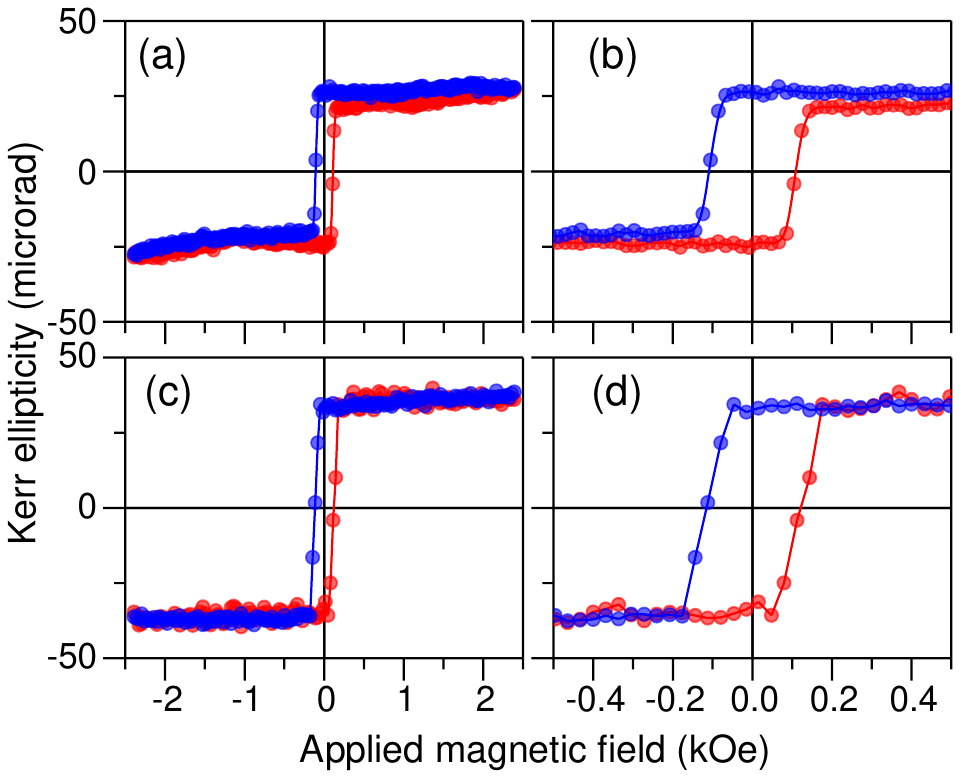}
	\caption{\label{Fig:Hystersisloops} The magneto-optical Kerr effect (MOKE) hysteresis loops recorded on  Co/Ni/Ir(001) [(a) and (b)] and 2Co/W(110) [(c) and (d)] samples. The loops were recorded in the longitudinal geometry at room temperature. In both cases the external magnetic field was applied along the [1$\bar{1}$0]-direction of the substrates. The red and blue colors indicate the upward and downward magnetic field sweep, respectively. Panels (b) and (d) represent a magnified part of panels (a) and (c), respectively.}
\end{figure}

\subsection{Film deposition and characterization}

Atomic layers of Co and Ni were deposited onto Ir(001) and W(110) at room temperature by means of electron-beam evaporation. The deposition rate for Ni and Co layers on Ir(001) was about  0.1 and 0.18 monolayer per minute (ML/min), respectively. In the case of Co films on W(110) the deposition rate was about 0.23 ML/min. 

The magnetic characterization of the samples was performed using magneto-optical Kerr effect (MOKE) in longitudinal geometry with the magnetic field applied along the [1$\bar{1}$0] direction of both the Ir(001) and W(110) substrates.  All the samples investigated in this study exhibit a ferromagnetic hysteresis loop as confirmed by our MOKE measurements. Typical hysteresis loops of the samples are shown in Fig.~\ref{Fig:Hystersisloops}, indicating a well-ordered magnetic state of the samples.

Prior to the spectroscopy experiments the sample was fully magnetized using a magnetic field of about $2.5$~kOe. The reason for such a high magnetic field is to avoid the presence of possible magnetic domains in the sample. Note that during the spectroscopy experiments no magnetic field was applied. It is, therefore, crucial to ensure that the sample is in the single domain state.
	
\subsection{Spin-polarized high-resolution electron energy-loss spectroscopy}
\begin{figure}[t]
	\centering
	\includegraphics[width=0.75\columnwidth]{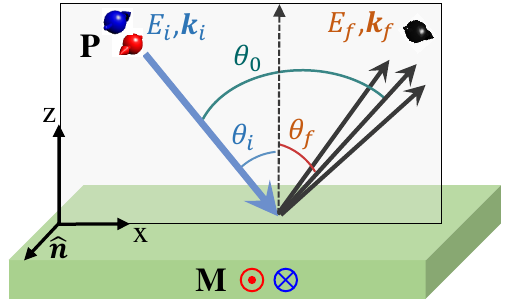}
	\caption{\label{Fig:ScatteringGeometry} A schematic representation of the scattering geometry
		used in the experiment. A spin-polarized beam with the polarization vector $\vec{P}$ is incident onto the sample surface. The direction of $\vec{P}$  is either parallel or antiparallel to the scattering's plane normal vector $\vec{\hat{n}}$. The
		incident  (scattered) beam energy and wavevector are denoted by $E_i$ ($E_f$) and $\vec{k_i}$ ($\vec{k_f}$), respectively. The magnon momentum is parallel to the $x$ axis and $z$ denotes the surface normal. The incident (scattered) beam angle is shown by $\theta_i$ ($\theta_f$). The total scattering angle is denoted by $\theta_0$. The shaded area represents the scattering plane and is given by the $x$ and $z$ axes.}
\end{figure}

SPHREELS was used to investigate the high-energy THz magnons. The technique has already been successfully implemented to excite exchange-dominated THz magnons at surfaces and in various ultrathin ferromagnetic films and multilayers  \cite{Vollmer2003,Zakeri2013, Zakeri2013b, Zakeri2014}. In addition, it can also address the spin-dependent electron scattering processes at the surfaces \cite{Kirschner1985,Zhang2011,Zakeri2023}. The scattering geometry used in the experiment is schematically shown in Fig.~\ref{Fig:ScatteringGeometry}. In SPHREELS experiments the magnitude of the magnon wavevector $\vec{Q}$ is given by $Q=Q_{\parallel}=k_{i\parallel}-k_{f\parallel}=k_{i}\sin(\theta_i)-k_{f}\sin(\theta_f)$ and is determined by the parallel component of the wavevector of the incident ($k_{i\parallel}$) an scattered ($k_{f\parallel}$) beam and the scattering angles ($\theta_i$ and $\theta_f$, see Fig.~\ref{Fig:ScatteringGeometry}). During the experiment the total scattering angle $\theta_0=\theta_i+\theta_f$ was set to 80$^{\circ}$. All the spectra were recorded at room temperature. In the experiment one may first fix the direction of the magnetization along a certain symmetry direction and measure the intensity profile of the scattered electrons for all the possible directions of the spin polarization vector of the incident beam $\vec{P}$. Of particular interest are two directions, i.e.,  parallel and antiparallel to the scattering plane's normal vector $\vec{\hat{n}}$. In case that the direction of the magnetization is also parallel and antiparallel to $\vec{\hat{n}}$ one can separate  the contributions of spin-orbit and exchange scattering to the observed spin asymmetry. Hence, in our experiments the scattering plane was adjusted to be along the [110] direction of the Ir(001) and [100] direction of W(110). The direction of $\vec{M}$ and $\vec{P}$ was either along [1$\bar{1}0$] or [$\bar{1}10$] (for both cases). Therefore, four different intensity spectra $I_{\mu \nu}$s were recorded for which $\mu$ represents the direction of $\vec{M}$, and $\nu$ that of $\vec{P}$. For instance, $I_{\uparrow \uparrow}$ ($I_{\downarrow \downarrow}$) indicates the scattered beam intensity when both $\vec{M}$ and $\vec{P}$ are parallel to [1$\bar{1}0$] ([$\bar{1}10$]). Likewise,  $I_{\uparrow \downarrow}$ ($I_{\downarrow \uparrow}$) represents a case for which $\vec{M}$ is  parallel to [1$\bar{1}0$]  ([$\bar{1}10$]) and $\vec{P}$ is parallel to [$\bar{1}10$] ([1$\bar{1}0$]).  

\begin{figure*}[t!]
	\centering
	\includegraphics[width=1.7\columnwidth]{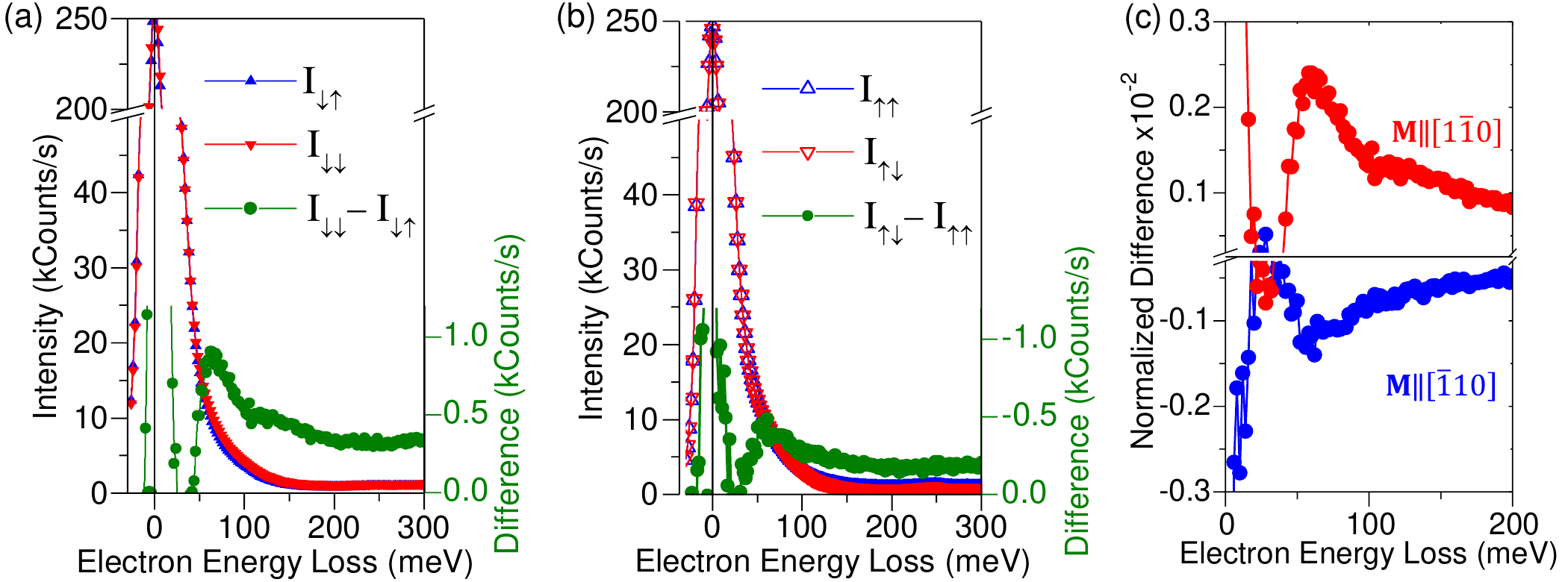}
	\caption{\label{Fig:SPEELS_Spectra} SPHREEL spectra recorded on the Co/Ni bilayer grown on Ir(001) at room temperature. The spectra were recorded at the wavevector of $|Q|=$0.4~\AA$^{-1}$ and the incident electron energy of $E_{i} = 10$~eV. The intensity spectra $I_{\mu \nu}$s represent the partial intensity spectra for a given direction of $\vec{M}$ and the spin-polarization vector $\vec{P}$, denoted by the subscripts $\mu$ and $\nu$, respectively. The spectra were recorded  for $\vec{M} \parallel [1\bar{1}0]$ in (a) and $\vec{M} \parallel [\bar{1}10]$ in (b). The difference spectra $I_{\mu \downarrow}-I_{\mu\uparrow}$ are shown in green. (c) The normalized difference spectra for the two directions of magnetization, i.e., $\vec{M} \parallel [1\bar{1}0]$ (red color) and $\vec{M} \parallel [\bar{1}10]$ (blue color).}
\end{figure*}

It is well-known that the magnon excitation process in SPHREELS is based on the exchange process \cite{Gokhale1992,Gokhale1994,Plihal1999,Zakeri2013}. Since a magnon carries a total angular momentum of $1\hbar$ it can only be excited by incidence of a minority electron (when $\vec{P}$ is parallel to $\vec{M}$). This means that in fact the difference spectra $I_{\mu \downarrow}-I_{\mu \uparrow}$, which describe a spin-flip process, shall contain all the information regarding the magnons as spin-flip excitations. Since the magnon excitation is mediated by the exchange process, in the absence of SOC the excitation amplitude should be independent from the direction of $\vec{M}$ with respect to the crystallographic directions. It should only depend on the relative orientation of $\vec{P}$ and $\vec{M}$. Hence, the intensity of the difference spectra is expected to be equal for the two magnetization directions, despite a negative sign. The presence of SOC during the scattering event leads to an inequality of the intensity of the difference spectra, since it couples the spin to the crystallographic directions. On the other hand reversing the direction of $\vec{M}$ is equivalent to reversing  the direction of the magnon wavevector \cite{Zakeri2010,Zakeri2012,Tsurkan2020}. This is because in a ferromagnet the time-reversal symmetry $\mathcal{T}$ is naturally broken. Hence, reversing the direction of $\vec{M}$ is equivalent to a time-reversal experiment. Therefore, the magnons for two opposite magnetization directions shall represent cases in which the magnons propagate along opposite directions. In principle, in the experiment changing the sign (or direction) of $\vec{Q}$ can be accomplished in two ways: (i) by changing the scattering geometry and (ii) by reversing the direction of $\vec{M}$ \cite{Zakeri2010,Zakeri2012}. Since changing the scattering geometry would change the scattering matrix elements, this may lead to unwanted effects. In order to avoid such effects one can simply reverse the direction of $\vec{M}$, keeping the scattering geometry unchanged (see Fig.~\ref{Fig:ScatteringGeometry}).
It has been shown that the presence of SOC leads to the fact that the magnon excitation amplitude, as seen by SPHREELS, is no longer equal for the two opposite directions of $\vec{M}$. The phenomenon is understood in terms of a strong competition between spin-orbit and exchange scattering mechanisms \cite{Zakeri2024}. 
While in most of ferromagnetic elements such as Fe, Co and Ni SOC is not significantly large, the spin-orbit scattering is expected to be small. However, in ultrathin films of ferromagnetic metals grown on substrates with a large SOC spin-orbit scattering can be significant. The presence of a large SOC can lead to an inequality of magnon amplitudes, a phenomenon known as magnon nonreciprocity. In order to understand the effect in detail we prepared and investigated two additional systems as will be discussed in the following section.

\section{Results and Discussion}

\subsection{The case of Co/Ni/Ir(001)}\label{Sec:Results_CoNi}

In order to address the impact of the interfacial electronic hybridization on the spin-orbit induced magnon nonreciprocity we started with the previously studied system, namely 2Co/Ir(001), and replaced the interface layer with a Ni layer. 
We investigated in detail the spin-dependence of both elastic and inelastic scattering with the particular attention on the magnon nonreciprocity. 

In Fig.~\ref{Fig:SPEELS_Spectra} the four possible partial intensity spectra recorded on the Co/Ni/Ir(001) system are presented. The spectra were recorded using an incident electron energy of $E_i=10$~eV and at a wavevector of $|Q|=0.4$~\AA$^{-1}$.  The difference spectra  $I_{\mu \downarrow}-I_{\mu \uparrow}$ shown in the green color in Figs.~\ref{Fig:SPEELS_Spectra}(a) and \ref{Fig:SPEELS_Spectra}(b) indicate a nonreciprocal magnon excitation in the present system. This is also clear by looking at the normalized difference spectra shown in Fig.~\ref{Fig:SPEELS_Spectra}(c). The normalization is done with respect to the total intensity of the quasielastic peak of each measurement, located at the energy loss of zero. The results indicate that the observed magnon nonreciprocity is smaller than that of the Co double layer grown on the same substrate \cite{Zakeri2024}.

\begin{figure}[t!]
	\centering
	\includegraphics[width=0.99\columnwidth]{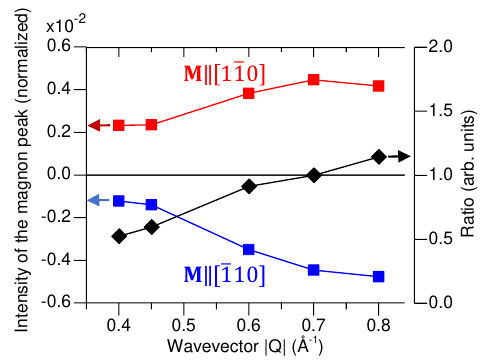}
	\caption{\label{Fig:Nonreciprocity} The intensity of the normalized difference spectra versus the norm of the magnon wavevector $|Q|$. The data were recorded for an incident electron energy of $E_{i} = 10$~eV and the two magnetization directions $\vec{M} \parallel [1\bar{1}0]$ (red) and $\vec{M} \parallel [\bar{1}10]$ (blue). The ratio of the two intensities is shown in back.}
\end{figure}

The observed magnon nonreciprocity is also present for magnons having other wavevectors (see the discussion below). The difference spectra presented in Fig.~\ref{Fig:SPEELS_Spectra} contain all the information regarding the excited magnon wavepackets, e.g., amplitude, lifetime, frequency, group and phase velocity, etc. One may, therefore, construct the magnon wavepackets propagating along two opposite directions based on these information \cite{Zhang2012,Qin2015}. Since the excitation energy (the magnon eigenfrequency), the lifetime, group and phase velocities are nearly the same for the two wavepackets, we realized that the main factor, which leads to the nonreciprocal propagation of the magnon wavepackets is due to the inequality of the amplitude of the wavepackets in the initial stage of propagation (just at the time when they are excited). It is important to notice that the nonreciprocal behavior of magnons can be an overall effect of the nonreciprocity in all characteristics of magnons, i.e., amplitude, frequency and lifetime. However, in the present case the nonreciprocity is mainly due to the nonreciprocal magnon amplitude. The nonreciprocity in the magnon frequency is rather small (on the order of 1~meV or less for this wavevector). This is due to the presence of the so-called chirality-inverted DMI. The effect has been discussed in detail in Refs.~\cite{Zakeri2023b,Zakeri2024b}. The amplitude of the magnon wavepackets is directly given by the height of the difference spectra at the magnon excitation energy. 
Hence, it would be very useful to analyze, in detail, the dependence of the ratio of the peak heights of magnon excitation spectra for  $\vec{M} \parallel [1\bar{1}0]$  and $\vec{M} \parallel [\bar{1}10]$ [see Fig.~\ref{Fig:SPEELS_Spectra}(c)]. This quantity may also be called as the magnon nonreciprocity. For a reciprocal propagation one expects a value equal to unity and for a unidirectional propagation this value should be zero. Since the absolute value of the intensity of the excitation spectra depends on the scattering geometry one should use the normalized difference spectra [similar to those shown in Fig.~\ref{Fig:SPEELS_Spectra}(c)]. In Fig.~\ref{Fig:Nonreciprocity} the excitation peak intensities for the two magnetization directions as well as their ratio are presented for different values of  $Q$. One immediate conclusion from the data presented in Fig.~\ref{Fig:Nonreciprocity} is that the magnon nonreciprocity depends strongly on $Q$ and can even be inverted for some ranges of $Q$. We may emphasize that the definition of the nonreciprocity is independent of the excitation method. It is valid for all excitation schemes, which involve the fundamental exchange process as the microscopic physical mechanism behind the magnon generation.

\begin{figure}[t!]
	\centering
	\includegraphics[width=0.97\columnwidth]{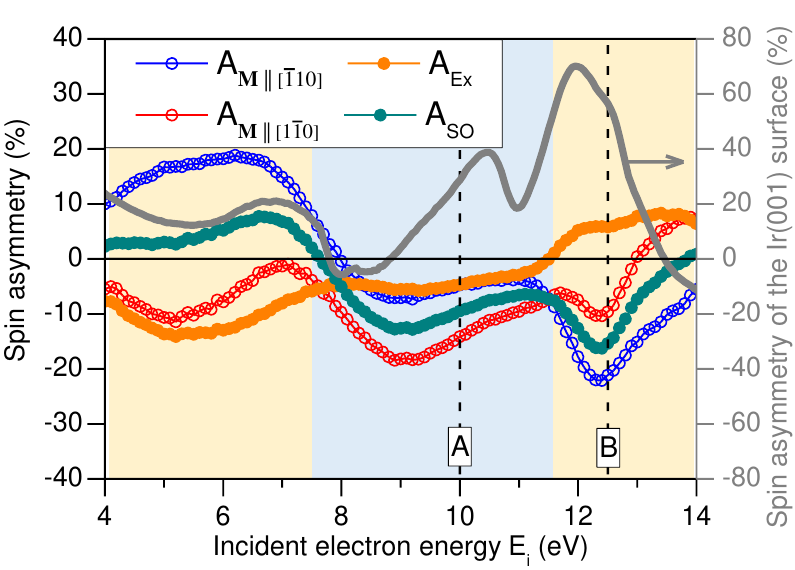}
	\caption{\label{Fig:Asymmetry_Energy_CoNi_Ir001} Spin asymmetries of elastically scattered electrons against the incident electron energy $E_i$. $A_{\vec{M} \parallel [1\bar{1}0]}$ and $A_{\vec{M} \parallel [\bar{1}10]}$ were recorded for two opposite directions of magnetization, $[1\bar{1}0]$ and $[\bar{1}10]$, respectively, as described in the text. Different contributions to the spin asymmetries are $A_{SO}$ and $A_{Ex}$ and are given by Eqs.~(\ref{Eq:SOA}) and (\ref{Eq:ExA}), respectively. The gray solid curve represents the spin-orbit asymmetry recorded on the bare Ir(001) surface. Note that the scale of the right axis is by a factor of 2 larger, compared to the left axis.}
\end{figure}

In order to understand the origin of this nonreciprocity, it is useful to separate the contributions of spin-orbit and exchange mechanisms to the scattering intensity. A precise separation is possible by performing a `complete' experiment in which a spin-polarized beam with a defined polarization vector is used and the spin polarization of the scattered electrons are measured after the scattering event \cite{Kirschner1985a}. Such an experiment would allow to measure all the possible eight partial intensity spectra $I_{\mu\nu}$s in which the direction of electrons spin $\nu$ is measured before and after the scattering event. In SPHREELS, due to the limitations associated with the high energy resolution and low count rates, the detection is not spin resolved \cite{Zakeri2014}. However, one can estimate the contributions of spin-orbit and exchange scattering to the asymmetry of the scattered beam intensity. To this end one may record and analyze the intensity of scattered electrons as a function of  $E_\mathrm{i}$ and $Q$ for two orientations of $\vec{M}$ and $\vec{P}$, e.g., parallel and antiparallel to $\vec{\hat{n}}$. The dependence of spin asymmetry on $E_i$ is shown in Fig.~\ref{Fig:Asymmetry_Energy_CoNi_Ir001} for both magnetization directions. The data were recorded for the specular geometry (elastic beam) which represents $Q=0$. The spin asymmetries are defined as $A_{\vec{M} \parallel [1\bar{1}0]} = (I_{\downarrow \downarrow}-I_{\downarrow \uparrow})/(I_{\downarrow \downarrow}+I_{\downarrow \uparrow})$ and $A_{\vec{M} \parallel [\bar{1}10]} = (I_{\uparrow \downarrow}-I_{\uparrow \uparrow})/(I_{\uparrow \downarrow}+I_{\uparrow \uparrow})$.
Assuming that the spin-orbit and exchange scattering events are two independent processes so that one can neglect the interference between them, the data would allow to separate the contribution of spin-orbit scattering and exchange scattering to the measured spin asymmetries \cite{Kirschner1985,Gradmann1986}. These are spin-orbit (SOA, $A_{SO}$) and exchange  (ExA, $A_{Ex}$) asymmetries.

While $A_{SO}$ is independent of the direction of $\vec{M}$, $A_{Ex}$ changes its sign when the direction of $\vec{M}$ is switched to the opposite direction.
Based on this simple argument and under the assumption mentioned above $A_{SO}$ and $A_{Ex}$ are  then given by the two following expressions.

\begin{multline}\label{Eq:SOA}
A_{SO} = \frac{I_{\downarrow \downarrow}-I_{\downarrow \uparrow}+I_{\uparrow \downarrow}-I_{\uparrow \uparrow}}{I_{\downarrow \downarrow}+I_{\downarrow \uparrow}+I_{\uparrow \downarrow}+I_{\uparrow \uparrow}}\\
\approx \frac{1}{2}\left(A_{\vec{M} \parallel [1\bar{1}0]}+A_{\vec{M} \parallel [\bar{1}10]}\right),
\end{multline}
  and
\begin{multline}\label{Eq:ExA}
	A_{Ex} = \frac{I_{\downarrow \downarrow}-I_{\downarrow \uparrow}-I_{\uparrow \downarrow}+I_{\uparrow \uparrow}}{I_{\downarrow \downarrow}+I_{\downarrow \uparrow}+I_{\uparrow \downarrow}+I_{\uparrow \uparrow}}\\
	\approx \frac{1}{2}\left(A_{\vec{M} \parallel [1\bar{1}0]}-A_{\vec{M} \parallel [\bar{1}10]}\right).
\end{multline}

Data shown in Fig.~\ref{Fig:Asymmetry_Energy_CoNi_Ir001} indicate that both SOA and ExA exhibit a strong dependence on $E_i$. The behavior is very similar to that of the Co double layer on the same substrate \cite{Zakeri2024}. In particular, the pattern of SOA is very similar in both systems, indicating that the SOC is mainly supported by the Ir(001) substrate. The $E_i$-dependence of SOA recorded on the bare Ir(001) substrate is also shown for a comparison (see the right axis of Fig.~\ref{Fig:Asymmetry_Energy_CoNi_Ir001}). It is important to notice that although the intrinsic SOC of the magnetic layers is not negligible and might be of some importance, it should be substantially enhanced in the presence of Ir as a heavy metal. Since  the interfacial electronic hybridization is expected to be marginal in the case of Ni/Ir interface, another important conclusion from this observation is that such a hybridization does not influence the behavior of SOA, drastically.  In order to carefully analyze the spin asymmetries one may consider several different regions of $E_i$. For instance, below 7.5~eV, $A_{SO}$ and $A_{Ex}$ possess a similar magnitude but an opposite sign (shown by shaded yellow color in Fig.~\ref{Fig:Asymmetry_Energy_CoNi_Ir001}). In the range of 7.5~eV$<E_i<11.5$~eV $A_{SO}$ and $A_{Ex}$  have the same sign but a different magnitude (light-blue shaded area in Fig.~\ref{Fig:Asymmetry_Energy_CoNi_Ir001}). For the incident energies above 11.5~eV $A_{SO}$ and $A_{Ex}$ exhibit again an opposite sign, similar to the first region (yellow shaded area in Fig.~\ref{Fig:Asymmetry_Energy_CoNi_Ir001}). In order to assess the consequences of the incident energy on the magnon nonreciprocity we analyzed the behavior of these asymmetries as a function of $Q$ for two areas marked by A and B in Fig.~\ref{Fig:Asymmetry_Energy_CoNi_Ir001}. Note that one could also analyze the data for $E_i<7.5$~eV. The results are expected to be similar to those of area B (see the data provided for the 2Co/W(110) system in Sec.~\ref{Sec:Results_2CoW}). 
Both SOA and ExA were measured as a function of $Q$ for two different values of the incident energy  (i) $E_{i}=10$~eV, located in area A, and (ii) $E_i=12.5$~eV, located in area B. Note that one can, in principle, choose any point within these areas as a representative point.

\begin{figure}[t!]
    \centering
    \includegraphics[width=0.95\columnwidth]{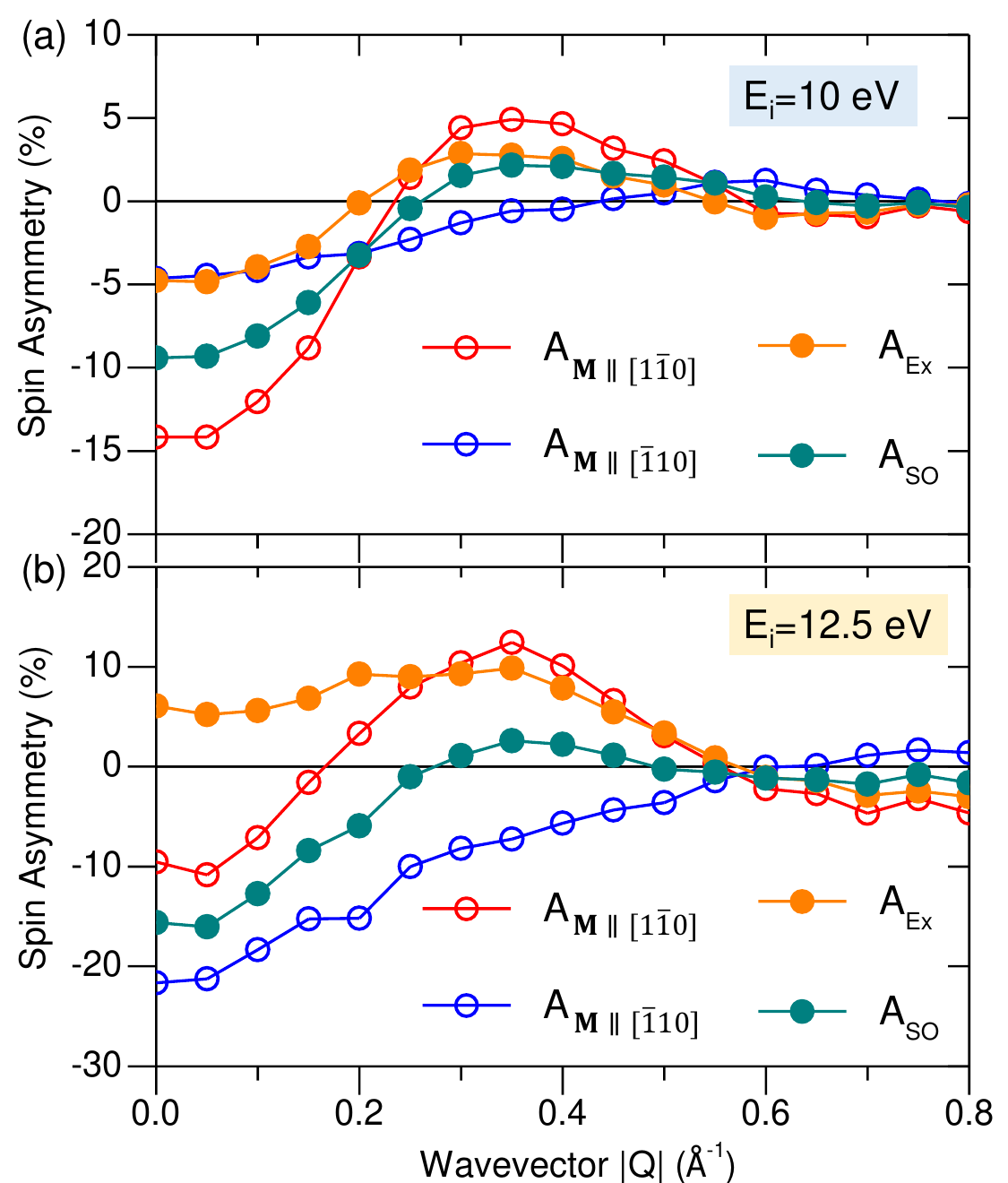}
    \caption{\label{Fig:Asymmetry_Wavevector_CoNi_Ir001} Spin asymmetries against the norm of the magnon wavevector $|Q|$. SOA and ExA are separated using Eqs.~(\ref{Eq:SOA}) and (\ref{Eq:ExA}), as discussed in the text. The results are shown for two values of incident energy (a) $E_{i} =$ 10~eV and (b) $E_{i} =$12.5~eV.}
\end{figure}

For the area A, where  SOA and ExA have the same sign but different magnitudes, we used $E_{i} =$10~eV as a representative energy  (marked as A in Fig.~\ref{Fig:Asymmetry_Energy_CoNi_Ir001}). This is the same incident energy used to record the energy-loss spectra presented in Fig.~\ref{Fig:SPEELS_Spectra}. The evolution of both SOA and ExA with $|Q|$ is presented in Fig.~\ref{Fig:Asymmetry_Wavevector_CoNi_Ir001}(a). The large variations of SOA  and ExA indicate the strong competition between the two scattering processes responsible for these asymmetries. In particular, in the vicinity of $|Q|=0.4$~\AA$^{-1}$ SOA is rather large and is comparable to ExA. Hence, it can influence the magnon excitation process. For one direction of magnetization, i.e., $[\overline{1}10]$ it would lead to a reduction of the magnon amplitude and for the opposite magnetization direction, i.e., $[1\overline{1}0]$  it would amplify the magnon amplitude. This is similar to the effect observed for the Co double layer on Ir(001) \cite{Zakeri2024}. The main difference is that the sign of ExA is positive in the present case.  
Note that the choice of $|Q|=0.4$~\AA$^{-1}$ is based on the fact that the magnon signal for this wavevector can easily be resolved. The effect is expected to be even larger at smaller magnon wavevectors.

Another result from the data presented in Fig.~\ref{Fig:Asymmetry_Wavevector_CoNi_Ir001}(a) is that SOA decreases for larger values of $Q$ and one would expect a small magnon nonreciprocity for $|Q| \approx 0.7$~\AA$^{-1}$. This is in agreement with the measurement of the magnon amplitude presented in Fig.~\ref{Fig:Nonreciprocity}. In a similar way the inversion of the nonreciprocity at  $|Q|=0.8$~\AA$^{-1}$ can also be explained based on the $Q$-dependence of spin asymmetries presented in Fig.~\ref{Fig:Asymmetry_Wavevector_CoNi_Ir001}(a).

 In order to investigate, in more detail, the $Q$-dependence of spin asymmetries for other incident energies, we moved to area B of Fig.~\ref{Fig:Asymmetry_Energy_CoNi_Ir001}, where SOA and ExA possess an opposite sign and a slightly different magnitude (SOA is slightly larger). We used $E_i=12.5$~eV as a representative point of this area and the results are shown in Fig.~\ref{Fig:Asymmetry_Wavevector_CoNi_Ir001}(b).
Again one observes a strong dependence of spin asymmetries on $|Q|$, which can lead to nonreciprocal excitation of magnons for some ranges of wavevector. However, SOA is considerably smaller than ExA (for example at $|Q|=0.4$~\AA$^{-1}$ it is smaller by a factor of about 3.5). This means that the expected nonreciprocal excitation of magnons for the incident energies within this area is small. Likewise, for  $E_i<7.5$~eV, where the exchange scattering mechanism is dominating (ExA is large and, at the same time, SOA is small, see Fig.~\ref{Fig:Asymmetry_Energy_CoNi_Ir001}), one would expect a small magnon nonreciprocity. One important point to consider is that the excitation cross-section of magnons by electrons depends critically on the incident beam energy. Hence, for different values of $E_i$ one would expect a different excitation cross-section. However, it is essential  to notice that the excitation cross-section should not depend on the direction of the magnetization as long as SOC is negligible.

We note that one may argue that the spin asymmetries presented in Figs.~\ref{Fig:Asymmetry_Energy_CoNi_Ir001} and \ref{Fig:Asymmetry_Wavevector_CoNi_Ir001} were measured for the elastically and quasielastically scattered electrons and not for the inelastically scattered ones. The magnon excitation peaks  appear at a certain energy-loss and in the inelastic part of the spectra.  However, one must pay attention to the following two important points. (i) If the spin-orbit scattering is significant in the elastic part of scattering, it should also be notable in the inelastic part \cite{Zakeri2022}. (ii) The magnon excitation is mediated by the exchange scattering processes, where an incident electron of minority character occupies an empty state above the Fermi-level and a majority electron from a state below the Fermi-level leaves the sample \cite{Kirschner1985a,Vollmer2003,Vollmer2004,Etzkorn2004,Zakeri2013,Zakeri2014}. The whole excitation process is, in fact, (quasi)elastic and the apparent energy loss of electron is due to the fact that the scattered electron leaves the sample from a state located at a lower energy. 
Based on the above mentioned arguments we conclude that the nonreciprocity of the magnon amplitude is a consequence of the competition between SOA and ExA, as it was also proposed in Ref.~\cite{Zakeri2024}. Both the sign and magnitude of SOA play a crucial role in this type of magnon nonreciprocity. 

As a side remark, it is important to emphasize that the magnon nonreciprocity is not directly a consequence of DMI, since it is observed to be large in systems with a small DMI-induced magnon nonreciprocity in the frequency (for instance the Co/Ir and Ni/Ir interfaces, which exhibit a chirality-inverted DMI \cite{Zakeri2023,Zakeri2024a,Zakeri2024b}). Moreover, the effect becomes more significant at lower wavevectors as seen in Figs.~\ref{Fig:Nonreciprocity} and \ref{Fig:Asymmetry_Wavevector_CoNi_Ir001} (see also Sec.~\ref{Sec:Conlusion}).

\begin{figure}[h!]
	\centering
	\includegraphics[width=0.85\columnwidth]{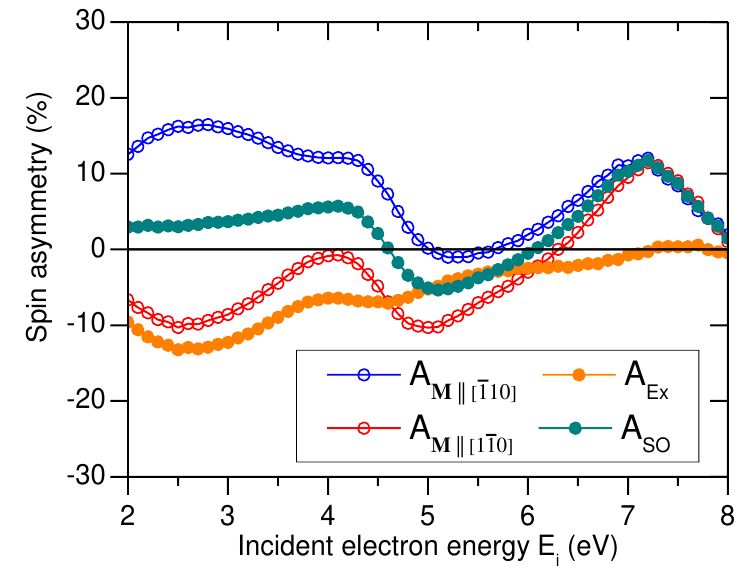}
	\caption{\label{Fig:Asymmetry_Ei_Co_W110} The same as Fig.~\ref{Fig:Asymmetry_Energy_CoNi_Ir001} but for the 2Co/W(110) system.}
\end{figure}

\subsection{The case of 2Co/W(110)}\label{Sec:Results_2CoW}

In a similar way we investigated a Co double layer on W(110). We first measured  the dependence of the spin-asymmetries on the incident beam energy $E_i$. The results are summarized in Fig.~\ref{Fig:Asymmetry_Ei_Co_W110}. As it was pointed out earlier, the magnon excitation cross-section by electrons as probing particles depends strongly on the incident electron energy. In the case of Co films on W(110) it was observed that the excitation cross-section is only notable in the vicinity of $E_i \approx 4$~eV and decreases drastically when changing the incident energy \cite{Etzkorn2004}. We, therefore, focused on this energy range and recorded the spin asymmetries in the close vicinity of $E_i = 4$~eV. The results for different  directions of magnetization are provided in Fig.~\ref{Fig:Asymmetry_Ei_Co_W110}. The contributions of SOA and ExA were separated using Eqs.~(\ref{Eq:SOA}) and (\ref{Eq:ExA}) and in a similar way as discussed in the case of Co/Ni/Ir(001) in Sec.~\ref{Sec:Results_CoNi}. 

The results indicate the presence of a large SOA in the measured range of $E_i$. However, within this range of incident energy ExA is also considerably large. The results are somewhat similar to those of Co double layer and Co/Ni bilayer on Ir(001). However, there are also several important differences. In particular, the change in the sign of SOA from positive to negative takes place at much lower values of $E_i$ in the present case ($E_i=4.5$ eV). Moreover, SOA changes once more the sign from negative to positive at about $E_i=6$~eV. 

\begin{figure}[b!]
	\centering
	\includegraphics[width=0.92\columnwidth]{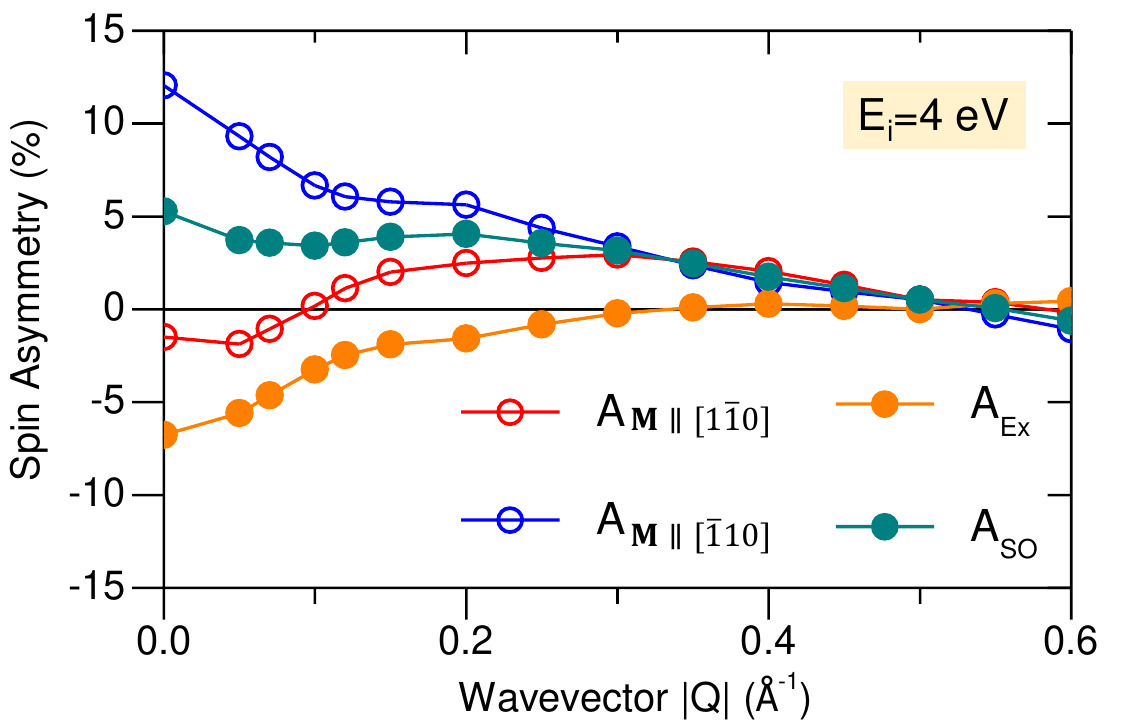}
	\caption{\label{Fig:Asymmetry_Wavevector_Co_W110} The same as Fig.~\ref{Fig:Asymmetry_Wavevector_CoNi_Ir001} but for the 2Co/W(110) system. The results were obtained for an incident energy of $E_i=4$~eV.}
\end{figure}

In order to clarify whether or not this large SOA can lead to any nonreciprocal excitation of magnons, we measured the $Q$-dependence of spin asymmetries at the incident energy of $E_i=4$ eV and the results are shown in Fig.~\ref{Fig:Asymmetry_Wavevector_Co_W110}. 
Similar to the results of the Co/Ni/Ir(001) system, in the present case one also observes that the spin asymmetries exhibit a strong $Q$-dependence. The results clearly demonstrate that one would also observe a magnon nonreciprocity for the present system as well. This is again a consequence of the spin-orbit scattering and its competition with the exchange mechanism. For the range of $Q$ accessible by the experiment  (consider the low incident energy of $E_i=4$~eV) the observed SOA and ExA are rather small, compared to those of the Co double layer and Co/Ni bilayer on Ir(001). However, SOA is still comparable to ExA and leads to a nonreciprocal magnon excitation. Hence, one would expect to observe a magnon nonreciprocity smaller compared to that of the magnetic systems grown on Ir(001). 

\begin{figure}[t!]
	\centering
	\includegraphics[width=1.0\columnwidth]{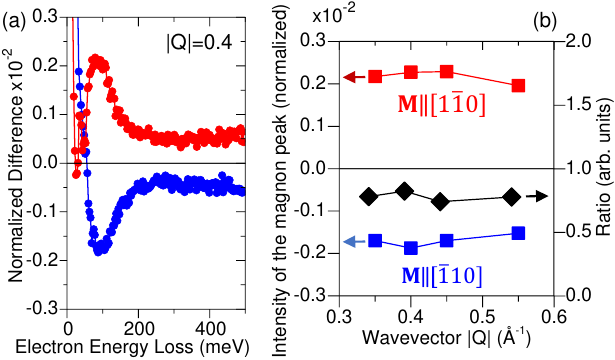}
	\caption{\label{Fig:Diff_2Co_W110}(a) Normalized difference spectra recorded for the two magnetization directions, i.e., [1$\bar{1}$0] and [$\bar{1}$10] at $|Q|=0.4$~\AA$^{-1}$. (b) The intensity of the normalized difference spectra for the two magnetization directions and their ratio plotted versus the norm of the magnon wavevector $|Q|$. The red (blue) color represent a case for which $ \vec{M}\parallel[1\bar{1}0]$ ($\vec{M}\parallel[\bar{1}10]$). All the results were obtained using an incident energy of $E_i=4$~eV on the 2Co/W(110) system.}
\end{figure}

In order to demonstrate this effect we measured the magnon spectra for the two magnetization directions. The results were obtained for different values of $|Q|$ ranging from 0.35 to 0.55 \AA$^{-1}$. As an example in Fig.~\ref{Fig:Diff_2Co_W110}(a) the normalized difference spectra recorded for $|Q|=0.4$~\AA$^{-1}$ are presented, indicating a nonzero magnon nonreciprocity. The dependence of the magnon nonreciprocity on $Q$ is demonstrated in Fig.~\ref{Fig:Diff_2Co_W110}(b), where both the normalized intensity of the magnon excitation peaks for the two magnetization directions as well as their ratio is plotted against $|Q|$. The results indicate that the magnon nonreciprocity is nearly constant within this range of wavevector. This is due to the fact that both SOA and ExA are nearly constant in this range.

Finally it is worth to comment on the presence of the antisymmetric DMI in the investigated systems. Similar to many other ultrathin ferromagnetic films grown on heavy-element substrates, DMI is also active in the present systems \cite{Zakeri2024b}. This fact should, in principle, lead to an asymmetric magnon dispersion relation and a nonreciprocity of the magnon energy (eigenfrequency)  \cite{Zakeri2010,Zakeri2012,Zakeri2017}. However, as discussed earlier in the  case of Co double layer and Co/Ni bilayer grown on Ir(001), due to the presence of a chirality-inverted DMI, the nonreciprocity in the magnon energy is rather small  \cite{Zakeri2023b,Zakeri2024b}. In the case of Co double layer on W(110) the DMI-induced magnon nonreciprocity in the energy is slightly larger \cite{Tsurkan2020}. However, the spin-orbit induced magnon nonreciprocity discussed here is a different phenomenon and appears in the magnon amplitude. This type of nonreciprocity becomes more significant at smaller magnon wavevectors (see Figs.~\ref{Fig:Nonreciprocity} and \ref{Fig:Asymmetry_Wavevector_CoNi_Ir001}).

\section{Concluding remarks} \label{Sec:Conlusion}

We investigated the spin-orbit induced magnon nonreciprocity in Co/Ni/Ir(001) and 2Co/W(110) and compared the results to those of the 2Co/Ir(001) system. The main aim was to investigate the influence of the electronic hybridization and the role of the substrate in the observed magnon nonreciprocity in detail. The results further confirm that a competition between the spin-orbit and exchange scattering leads to the magnon nonreciprocity, as it was proposed in Ref.~\cite{Zakeri2024}. 

A comparison between the results of Co/Ni/Ir(001) to those of 2Co/Ir(001) indicates that despite of the fact that for Co/Ni/Ir(001) the magnon nonreciprocity is smaller, the main features are very similar for the two systems. This indicates that although the interfacial hybridization of the electronic states is somewhat important, it is not decisive for the overall behavior of the magnon nonreciprocity. It was observed that since both SOA and ExA depend strongly on the energy of the incident beam and the scattering geometry, the magnon nonreciprocity depends also on these variables. The nonreciprocity can even be inverted for some ranges of the incident energy $E_i$ and magnon wavevector $Q$. 

In the case of the Co double layer on W(110) the magnon nonreciprocity was found to be less significant compared to that observed for the  layered structures grown on Ir(001). The effect could also be explained based on the competition between SOA and ExA. Comparing the results to those of the Co/Ni bilayer and Co double layer on Ir(001) indicates that the choice of the substrate is important for the observed effect. The substrate influences the spin-dependent electron scattering through a combination of the surface symmetry and the spin-orbit interaction.

The fact that the magnon nonreciprocity for the case of 2Co/W(110) presented in Fig.~\ref{Fig:Diff_2Co_W110} is smaller (compared to that of the 2Co/Ir(001) and Co/Ni/Ir(001)) is due to the fact that the data for  2Co/W(110) are presented for $E_i=4$~eV, where ExA is larger than SOA (see Fig.~\ref{Fig:Asymmetry_Ei_Co_W110}). Moreover, as seen in Fig.~\ref{Fig:Asymmetry_Wavevector_Co_W110} at these wavevectors SOA is rather small. For the case of the Co/Ni/Ir(001) system the magnon nonreciprocity is presented for the incident energy of $E_i=10$~eV, where SOA is larger than ExA. This is also the case when looking at the $Q$-dependence of SOA, presented in Fig.~\ref{Fig:Asymmetry_Wavevector_CoNi_Ir001}. 

The results clearly demonstrate that both the incident beam energy and the choice of the substrate provide ways to tune the magnon nonreciprocity. This may inspire novel ideas for excitation of nonreciprocal ultrafast THz magnons for the application in the field of magnonics. We anticipate that the effect discussed here is not restricted to electron scattering experiments. It should also be observed in transport-based excitation schemes \cite{Tsui1971,Moodera1998,Murai1999,Fuchs2004}. 

Moreover, it is not limited to ultrathin films of $3d$ ferromagnets grown on metallic substrates. A similar effect is also expected when the underlying substrate is not metallic , e.g., dielectric oxides \cite{Zakeri2023a}. As long as the spin-orbit scattering is large one should be able to observe a magnon nonreciprocity. Analogously, the same effect is also expected for two-dimensional (2D) van der Waals magnets as well as heterostructures based on such 2D materials. One idea is to fabricate a bilayer made of a 2D ferromagnet and another 2D material which possesses a large SOC. Likewise, one may think of a 2D ferromagnet put on top of a topological insulator, e.g., Bi$_2$Se$_3$(0001) \cite{Zakeri2021b}. Since the interfacial electronic hybridization is not decisive for this phenomenon, one should, in principle, observe a notable magnon nonreciprocity.

In addition to ultrathin film systems discussed here, we anticipate that the phenomenon should also be present in other combinations of $3d$ magnetic elements and heavy-element substrates e.g., Pt. Other candidates are, the $3d$--$5d$ alloys, e.g., FePt and CoPt or even more complex alloys, e.g.,  magnetic iridates and ruthenates, which possess a large SOC.

In principle, the effect shall also be present for systems with a moderate SOC. Since it scales with the strength of SOC, it can be best observed in the system in which this interaction is strong. Due to the complex nature of inelastic electron scattering it is not straightforward to estimate the expected magnon nonreciprocity for any arbitrary system. However, as a rule of thumb we expect that the effect would at least be proportional to the strength of SOC.

Another consideration is that our experiments were performed on high-quality epitaxial samples. In such a case the in-plane magnon momentum is well-defined and one can follow the dispersion relation of magnons with a high degree of accuracy. In the case of polycrystalline samples the magnon momentum is no longer a well-defined quantity and one deals with the magnons integrated over a certain range of momentum, depending on the degree of epitaxy. In order to observe the effect discussed here one requires a certain degree of epitaxy to avoid cancellation effects.

Since the magnon nonreciprocity is mainly caused by the fact that the excitation cross-section of magnons is different for the two opposite propagation directions (or the two opposite directions of $\vec{M}$), it is directly linked to the exchange process as the microscopic mechanism behind the magnon excitation. It is, therefore, expected that this kind of magnon nonreciprocity should be observed in all excitation schemes, which involve the exchange mechanism as the underlying mechanism for magnon excitation. For instance, in the transport-based tunneling experiments, in which a spin-polarized current is used to excite magnons e.g., in tunneling magnetoresistance type of devices (see for example \cite{Tsui1971,Moodera1998,Murai1999,Fuchs2004,Balashov2006,Dzyapko2013,Ghazaryan2018}). Likewise, in the other excitation schemes, in which spin-polarized electrons are injected into a ferromagnetic layer, the observed magnon nonreciprocity should be present \cite{Razdolski2017}. The effect should also be observed in other scattering methods based on inelastic neutrons and photons, as the amplitude of magnon wavepackets for opposite propagation directions (or the two opposite directions of $\vec{M}$) is intrinsically different.

If the aim is to utilize the effect in practical applications one needs to think of possible obstacles and possible strategies to overcome such obstacles. The feasibility of using THz magnons for information processing has been discussed in detail in Refs.~\cite{Zakeri2018,Zakeri2020}.
It is worth mentioning that one of the main advantages of these excitations is that they are ultrafast and are confined on nanometer  length scales. Hence, they would allow for  fabrication of nanoscale devices \cite{Zakeri2024c}.

\begin{acknowledgments}

The financial support of the Deutsche Forschungsgemeinschaft (DFG) through the DFG Grants ZA 902/5-1, ZA 902/7-1 and ZA 902/8-1 is acknowledged. We would like to thank the Physikalisches Institut for hosting the group and for providing the infrastructure.

\end{acknowledgments}

\bibliography{Refs}

\begin{thebibliography}{70}%
\makeatletter
\providecommand \@ifxundefined [1]{%
 \@ifx{#1\undefined}
}%
\providecommand \@ifnum [1]{%
 \ifnum #1\expandafter \@firstoftwo
 \else \expandafter \@secondoftwo
 \fi
}%
\providecommand \@ifx [1]{%
 \ifx #1\expandafter \@firstoftwo
 \else \expandafter \@secondoftwo
 \fi
}%
\providecommand \natexlab [1]{#1}%
\providecommand \enquote  [1]{``#1''}%
\providecommand \bibnamefont  [1]{#1}%
\providecommand \bibfnamefont [1]{#1}%
\providecommand \citenamefont [1]{#1}%
\providecommand \href@noop [0]{\@secondoftwo}%
\providecommand \href [0]{\begingroup \@sanitize@url \@href}%
\providecommand \@href[1]{\@@startlink{#1}\@@href}%
\providecommand \@@href[1]{\endgroup#1\@@endlink}%
\providecommand \@sanitize@url [0]{\catcode `\\12\catcode `\$12\catcode
  `\&12\catcode `\#12\catcode `\^12\catcode `\_12\catcode `\%12\relax}%
\providecommand \@@startlink[1]{}%
\providecommand \@@endlink[0]{}%
\providecommand \url  [0]{\begingroup\@sanitize@url \@url }%
\providecommand \@url [1]{\endgroup\@href {#1}{\urlprefix }}%
\providecommand \urlprefix  [0]{URL }%
\providecommand \Eprint [0]{\href }%
\providecommand \doibase [0]{https://doi.org/}%
\providecommand \selectlanguage [0]{\@gobble}%
\providecommand \bibinfo  [0]{\@secondoftwo}%
\providecommand \bibfield  [0]{\@secondoftwo}%
\providecommand \translation [1]{[#1]}%
\providecommand \BibitemOpen [0]{}%
\providecommand \bibitemStop [0]{}%
\providecommand \bibitemNoStop [0]{.\EOS\space}%
\providecommand \EOS [0]{\spacefactor3000\relax}%
\providecommand \BibitemShut  [1]{\csname bibitem#1\endcsname}%
\let\auto@bib@innerbib\@empty
\bibitem [{\citenamefont {Feng}\ \emph {et~al.}(2011)\citenamefont {Feng},
  \citenamefont {Ayache}, \citenamefont {Huang}, \citenamefont {Xu},
  \citenamefont {Lu}, \citenamefont {Chen}, \citenamefont {Fainman},\ and\
  \citenamefont {Scherer}}]{Feng2011}%
  \BibitemOpen
  \bibfield  {author} {\bibinfo {author} {\bibfnamefont {L.}~\bibnamefont
  {Feng}}, \bibinfo {author} {\bibfnamefont {M.}~\bibnamefont {Ayache}},
  \bibinfo {author} {\bibfnamefont {J.}~\bibnamefont {Huang}}, \bibinfo
  {author} {\bibfnamefont {Y.-L.}\ \bibnamefont {Xu}}, \bibinfo {author}
  {\bibfnamefont {M.-H.}\ \bibnamefont {Lu}}, \bibinfo {author} {\bibfnamefont
  {Y.-F.}\ \bibnamefont {Chen}}, \bibinfo {author} {\bibfnamefont
  {Y.}~\bibnamefont {Fainman}},\ and\ \bibinfo {author} {\bibfnamefont
  {A.}~\bibnamefont {Scherer}},\ }\bibfield  {title} {\bibinfo {title}
  {Nonreciprocal light propagation in a silicon photonic circuit},\ }\href
  {https://doi.org/10.1126/science.1206038} {\bibfield  {journal} {\bibinfo
  {journal} {Science}\ }\textbf {\bibinfo {volume} {333}},\ \bibinfo {pages}
  {729} (\bibinfo {year} {2011})}\BibitemShut {NoStop}%
\bibitem [{\citenamefont {Ramezani}\ \emph {et~al.}(2018)\citenamefont
  {Ramezani}, \citenamefont {Jha}, \citenamefont {Wang},\ and\ \citenamefont
  {Zhang}}]{Ramezani2018}%
  \BibitemOpen
  \bibfield  {author} {\bibinfo {author} {\bibfnamefont {H.}~\bibnamefont
  {Ramezani}}, \bibinfo {author} {\bibfnamefont {P.~K.}\ \bibnamefont {Jha}},
  \bibinfo {author} {\bibfnamefont {Y.}~\bibnamefont {Wang}},\ and\ \bibinfo
  {author} {\bibfnamefont {X.}~\bibnamefont {Zhang}},\ }\bibfield  {title}
  {\bibinfo {title} {Nonreciprocal localization of photons},\ }\href
  {https://doi.org/10.1103/physrevlett.120.043901} {\bibfield  {journal}
  {\bibinfo  {journal} {Physical Review Letters}\ }\textbf {\bibinfo {volume}
  {120}},\ \bibinfo {pages} {043901} (\bibinfo {year} {2018})}\BibitemShut
  {NoStop}%
\bibitem [{\citenamefont {Damon}\ and\ \citenamefont
  {Eshbach}(1961)}]{Damon1961}%
  \BibitemOpen
  \bibfield  {author} {\bibinfo {author} {\bibfnamefont {R.~W.}\ \bibnamefont
  {Damon}}\ and\ \bibinfo {author} {\bibfnamefont {J.~R.}\ \bibnamefont
  {Eshbach}},\ }\bibfield  {title} {\bibinfo {title} {Magnetostatic modes of a
  ferromagnet slab},\ }\href {https://doi.org/10.1016/0022-3697(61)90041-5}
  {\bibfield  {journal} {\bibinfo  {journal} {Journal of Physics and Chemistry
  of Solids}\ }\textbf {\bibinfo {volume} {19}},\ \bibinfo {pages} {308}
  (\bibinfo {year} {1961})}\BibitemShut {NoStop}%
\bibitem [{\citenamefont {Gr\"unberg}(1985)}]{Gruenberg1985}%
  \BibitemOpen
  \bibfield  {author} {\bibinfo {author} {\bibfnamefont {P.}~\bibnamefont
  {Gr\"unberg}},\ }\bibfield  {title} {\bibinfo {title} {Light scattering from
  magnetic surface excitations},\ }\href
  {https://doi.org/10.1016/0079-6816(85)90009-7} {\bibfield  {journal}
  {\bibinfo  {journal} {Progress in Surface Science}\ }\textbf {\bibinfo
  {volume} {18}},\ \bibinfo {pages} {1 } (\bibinfo {year} {1985})}\BibitemShut
  {NoStop}%
\bibitem [{\citenamefont {Seki}\ \emph {et~al.}(2016)\citenamefont {Seki},
  \citenamefont {Okamura}, \citenamefont {Kondou}, \citenamefont {Shibata},
  \citenamefont {Kubota}, \citenamefont {Takagi}, \citenamefont {Kagawa},
  \citenamefont {Kawasaki}, \citenamefont {Tatara}, \citenamefont {Otani},\
  and\ \citenamefont {Tokura}}]{Seki2016}%
  \BibitemOpen
  \bibfield  {author} {\bibinfo {author} {\bibfnamefont {S.}~\bibnamefont
  {Seki}}, \bibinfo {author} {\bibfnamefont {Y.}~\bibnamefont {Okamura}},
  \bibinfo {author} {\bibfnamefont {K.}~\bibnamefont {Kondou}}, \bibinfo
  {author} {\bibfnamefont {K.}~\bibnamefont {Shibata}}, \bibinfo {author}
  {\bibfnamefont {M.}~\bibnamefont {Kubota}}, \bibinfo {author} {\bibfnamefont
  {R.}~\bibnamefont {Takagi}}, \bibinfo {author} {\bibfnamefont
  {F.}~\bibnamefont {Kagawa}}, \bibinfo {author} {\bibfnamefont
  {M.}~\bibnamefont {Kawasaki}}, \bibinfo {author} {\bibfnamefont
  {G.}~\bibnamefont {Tatara}}, \bibinfo {author} {\bibfnamefont
  {Y.}~\bibnamefont {Otani}},\ and\ \bibinfo {author} {\bibfnamefont
  {Y.}~\bibnamefont {Tokura}},\ }\bibfield  {title} {\bibinfo {title}
  {Magnetochiral nonreciprocity of volume spin wave propagation in
  chiral-lattice ferromagnets},\ }\href
  {https://doi.org/10.1103/PhysRevB.93.235131} {\bibfield  {journal} {\bibinfo
  {journal} {Phys. Rev. B}\ }\textbf {\bibinfo {volume} {93}},\ \bibinfo
  {pages} {235131} (\bibinfo {year} {2016})}\BibitemShut {NoStop}%
\bibitem [{\citenamefont {Mruczkiewicz}\ \emph {et~al.}(2017)\citenamefont
  {Mruczkiewicz}, \citenamefont {Graczyk}, \citenamefont {Lupo}, \citenamefont
  {Adeyeye}, \citenamefont {Gubbiotti},\ and\ \citenamefont
  {Krawczyk}}]{Mruczkiewicz2017}%
  \BibitemOpen
  \bibfield  {author} {\bibinfo {author} {\bibfnamefont {M.}~\bibnamefont
  {Mruczkiewicz}}, \bibinfo {author} {\bibfnamefont {P.}~\bibnamefont
  {Graczyk}}, \bibinfo {author} {\bibfnamefont {P.}~\bibnamefont {Lupo}},
  \bibinfo {author} {\bibfnamefont {A.}~\bibnamefont {Adeyeye}}, \bibinfo
  {author} {\bibfnamefont {G.}~\bibnamefont {Gubbiotti}},\ and\ \bibinfo
  {author} {\bibfnamefont {M.}~\bibnamefont {Krawczyk}},\ }\bibfield  {title}
  {\bibinfo {title} {Spin-wave nonreciprocity and magnonic band structure in a
  thin permalloy film induced by dynamical coupling with an array of {N}i
  stripes},\ }\href {https://doi.org/10.1103/PhysRevB.96.104411} {\bibfield
  {journal} {\bibinfo  {journal} {Physical Review B}\ }\textbf {\bibinfo
  {volume} {96}},\ \bibinfo {pages} {104411} (\bibinfo {year}
  {2017})}\BibitemShut {NoStop}%
\bibitem [{\citenamefont {Ishibashi}\ \emph {et~al.}(2020)\citenamefont
  {Ishibashi}, \citenamefont {Shiota}, \citenamefont {Li}, \citenamefont
  {Funada}, \citenamefont {Moriyama},\ and\ \citenamefont
  {Ono}}]{Ishibashi2020}%
  \BibitemOpen
  \bibfield  {author} {\bibinfo {author} {\bibfnamefont {M.}~\bibnamefont
  {Ishibashi}}, \bibinfo {author} {\bibfnamefont {Y.}~\bibnamefont {Shiota}},
  \bibinfo {author} {\bibfnamefont {T.}~\bibnamefont {Li}}, \bibinfo {author}
  {\bibfnamefont {S.}~\bibnamefont {Funada}}, \bibinfo {author} {\bibfnamefont
  {T.}~\bibnamefont {Moriyama}},\ and\ \bibinfo {author} {\bibfnamefont
  {T.}~\bibnamefont {Ono}},\ }\bibfield  {title} {\bibinfo {title} {Switchable
  giant nonreciprocal frequency shift of propagating spin waves in synthetic
  antiferromagnets},\ }\href {https://doi.org/10.1126/sciadv.aaz6931}
  {\bibfield  {journal} {\bibinfo  {journal} {Science Advances}\ }\textbf
  {\bibinfo {volume} {6}},\ \bibinfo {pages} {eaaz6931} (\bibinfo {year}
  {2020})}\BibitemShut {NoStop}%
\bibitem [{\citenamefont {Wang}\ \emph {et~al.}(2020)\citenamefont {Wang},
  \citenamefont {Kewenig}, \citenamefont {Schneider}, \citenamefont {Verba},
  \citenamefont {Kohl}, \citenamefont {Heinz}, \citenamefont {Geilen},
  \citenamefont {Mohseni}, \citenamefont {L\"agel}, \citenamefont {Ciubotaru},
  \citenamefont {Adelmann}, \citenamefont {Dubs}, \citenamefont {Cotofana},
  \citenamefont {Dobrovolskiy}, \citenamefont {Br\"acher}, \citenamefont
  {Pirro},\ and\ \citenamefont {Chumak}}]{Wang2020}%
  \BibitemOpen
  \bibfield  {author} {\bibinfo {author} {\bibfnamefont {Q.}~\bibnamefont
  {Wang}}, \bibinfo {author} {\bibfnamefont {M.}~\bibnamefont {Kewenig}},
  \bibinfo {author} {\bibfnamefont {M.}~\bibnamefont {Schneider}}, \bibinfo
  {author} {\bibfnamefont {R.}~\bibnamefont {Verba}}, \bibinfo {author}
  {\bibfnamefont {F.}~\bibnamefont {Kohl}}, \bibinfo {author} {\bibfnamefont
  {B.}~\bibnamefont {Heinz}}, \bibinfo {author} {\bibfnamefont
  {M.}~\bibnamefont {Geilen}}, \bibinfo {author} {\bibfnamefont
  {M.}~\bibnamefont {Mohseni}}, \bibinfo {author} {\bibfnamefont
  {B.}~\bibnamefont {L\"agel}}, \bibinfo {author} {\bibfnamefont
  {F.}~\bibnamefont {Ciubotaru}}, \bibinfo {author} {\bibfnamefont
  {C.}~\bibnamefont {Adelmann}}, \bibinfo {author} {\bibfnamefont
  {C.}~\bibnamefont {Dubs}}, \bibinfo {author} {\bibfnamefont {S.~D.}\
  \bibnamefont {Cotofana}}, \bibinfo {author} {\bibfnamefont {O.~V.}\
  \bibnamefont {Dobrovolskiy}}, \bibinfo {author} {\bibfnamefont
  {T.}~\bibnamefont {Br\"acher}}, \bibinfo {author} {\bibfnamefont
  {P.}~\bibnamefont {Pirro}},\ and\ \bibinfo {author} {\bibfnamefont {A.~V.}\
  \bibnamefont {Chumak}},\ }\bibfield  {title} {\bibinfo {title} {A magnonic
  directional coupler for integrated magnonic half-adders},\ }\href
  {https://doi.org/10.1038/s41928-020-00485-6} {\bibfield  {journal} {\bibinfo
  {journal} {Nature Electronics}\ }\textbf {\bibinfo {volume} {3}},\ \bibinfo
  {pages} {765} (\bibinfo {year} {2020})}\BibitemShut {NoStop}%
\bibitem [{\citenamefont {Dobrovolskiy}\ and\ \citenamefont
  {Chumak}(2022)}]{Dobrovolskiy2022}%
  \BibitemOpen
  \bibfield  {author} {\bibinfo {author} {\bibfnamefont {O.~V.}\ \bibnamefont
  {Dobrovolskiy}}\ and\ \bibinfo {author} {\bibfnamefont {A.~V.}\ \bibnamefont
  {Chumak}},\ }\bibfield  {title} {\bibinfo {title} {Nonreciprocal magnon
  fluxonics upon ferromagnet/superconductor hybrids},\ }\href
  {https://doi.org/10.1016/j.jmmm.2021.168633} {\bibfield  {journal} {\bibinfo
  {journal} {J. Magn. Magn. Mater.}\ }\textbf {\bibinfo {volume} {543}},\
  \bibinfo {pages} {168633} (\bibinfo {year} {2022})}\BibitemShut {NoStop}%
\bibitem [{\citenamefont {Gladii}\ \emph {et~al.}(2023)\citenamefont {Gladii},
  \citenamefont {Salikhov}, \citenamefont {Hellwig}, \citenamefont
  {Schultheiss}, \citenamefont {Lindner},\ and\ \citenamefont
  {Gallardo}}]{Gladii2023}%
  \BibitemOpen
  \bibfield  {author} {\bibinfo {author} {\bibfnamefont {O.}~\bibnamefont
  {Gladii}}, \bibinfo {author} {\bibfnamefont {R.}~\bibnamefont {Salikhov}},
  \bibinfo {author} {\bibfnamefont {O.}~\bibnamefont {Hellwig}}, \bibinfo
  {author} {\bibfnamefont {H.}~\bibnamefont {Schultheiss}}, \bibinfo {author}
  {\bibfnamefont {J.}~\bibnamefont {Lindner}},\ and\ \bibinfo {author}
  {\bibfnamefont {R.~A.}\ \bibnamefont {Gallardo}},\ }\bibfield  {title}
  {\bibinfo {title} {Spin-wave nonreciprocity at the spin-flop transition
  region in synthetic antiferromagnets},\ }\href
  {https://doi.org/10.1103/physrevb.107.104419} {\bibfield  {journal} {\bibinfo
   {journal} {Physical Review B}\ }\textbf {\bibinfo {volume} {107}},\ \bibinfo
  {pages} {104419} (\bibinfo {year} {2023})}\BibitemShut {NoStop}%
\bibitem [{\citenamefont {Mills}(2007)}]{Mills2007}%
  \BibitemOpen
  \bibfield  {author} {\bibinfo {author} {\bibfnamefont {D.~L.}\ \bibnamefont
  {Mills}},\ }\bibinfo {title} {Spin waves: {H}istory and a summary of recent
  developments}\ (\bibinfo  {publisher} {Wiley \& Sons, Ltd.},\ \bibinfo {year}
  {2007})\ Chap.\ \bibinfo {chapter} {Fundamentals and Theory}, pp.\ \bibinfo
  {pages} {247--282}\BibitemShut {NoStop}%
\bibitem [{\citenamefont {Zakeri}(2014)}]{Zakeri2014}%
  \BibitemOpen
  \bibfield  {author} {\bibinfo {author} {\bibfnamefont {K.}~\bibnamefont
  {Zakeri}},\ }\bibfield  {title} {\bibinfo {title} {Elementary spin
  excitations in ultrathin itinerant magnets},\ }\href
  {https://doi.org/h10.1016/j.physrep.2014.08.001} {\bibfield  {journal}
  {\bibinfo  {journal} {Phys. Rep.}\ }\textbf {\bibinfo {volume} {545}},\
  \bibinfo {pages} {47} (\bibinfo {year} {2014})}\BibitemShut {NoStop}%
\bibitem [{\citenamefont {Raghu}\ \emph {et~al.}(2010)\citenamefont {Raghu},
  \citenamefont {Chung}, \citenamefont {Qi},\ and\ \citenamefont
  {Zhang}}]{Raghu2010}%
  \BibitemOpen
  \bibfield  {author} {\bibinfo {author} {\bibfnamefont {S.}~\bibnamefont
  {Raghu}}, \bibinfo {author} {\bibfnamefont {S.~B.}\ \bibnamefont {Chung}},
  \bibinfo {author} {\bibfnamefont {X.-L.}\ \bibnamefont {Qi}},\ and\ \bibinfo
  {author} {\bibfnamefont {S.-C.}\ \bibnamefont {Zhang}},\ }\bibfield  {title}
  {\bibinfo {title} {Collective modes of a helical liquid},\ }\href
  {https://doi.org/10.1103/physrevlett.104.116401} {\bibfield  {journal}
  {\bibinfo  {journal} {Physical Review Letters}\ }\textbf {\bibinfo {volume}
  {104}},\ \bibinfo {pages} {116401} (\bibinfo {year} {2010})}\BibitemShut
  {NoStop}%
\bibitem [{\citenamefont {Zakeri}\ \emph
  {et~al.}(2021{\natexlab{a}})\citenamefont {Zakeri}, \citenamefont
  {Wettstein},\ and\ \citenamefont {S\"urgers}}]{Zakeri2021b}%
  \BibitemOpen
  \bibfield  {author} {\bibinfo {author} {\bibfnamefont {K.}~\bibnamefont
  {Zakeri}}, \bibinfo {author} {\bibfnamefont {J.}~\bibnamefont {Wettstein}},\
  and\ \bibinfo {author} {\bibfnamefont {C.}~\bibnamefont {S\"urgers}},\
  }\bibfield  {title} {\bibinfo {title} {Generation of spin-polarized hot
  electrons at topological insulators surfaces by scattering from collective
  charge excitations},\ }\href {https://doi.org/10.1038/s42005-021-00729-7}
  {\bibfield  {journal} {\bibinfo  {journal} {Commun. Phys.}\ }\textbf
  {\bibinfo {volume} {4}},\ \bibinfo {pages} {225} (\bibinfo {year}
  {2021}{\natexlab{a}})}\BibitemShut {NoStop}%
\bibitem [{\citenamefont {Zakeri}\ \emph
  {et~al.}(2010{\natexlab{a}})\citenamefont {Zakeri}, \citenamefont {Zhang},
  \citenamefont {Prokop}, \citenamefont {Chuang}, \citenamefont {Sakr},
  \citenamefont {Tang},\ and\ \citenamefont {Kirschner}}]{Zakeri2010}%
  \BibitemOpen
  \bibfield  {author} {\bibinfo {author} {\bibfnamefont {K.}~\bibnamefont
  {Zakeri}}, \bibinfo {author} {\bibfnamefont {Y.}~\bibnamefont {Zhang}},
  \bibinfo {author} {\bibfnamefont {J.}~\bibnamefont {Prokop}}, \bibinfo
  {author} {\bibfnamefont {T.-H.}\ \bibnamefont {Chuang}}, \bibinfo {author}
  {\bibfnamefont {N.}~\bibnamefont {Sakr}}, \bibinfo {author} {\bibfnamefont
  {W.~X.}\ \bibnamefont {Tang}},\ and\ \bibinfo {author} {\bibfnamefont
  {J.}~\bibnamefont {Kirschner}},\ }\bibfield  {title} {\bibinfo {title}
  {Asymmetric spin-wave dispersion on {F}e(110): {D}irect evidence of the
  {D}zyaloshinskii-{M}oriya interaction},\ }\href
  {https://doi.org/10.1103/PhysRevLett.104.137203} {\bibfield  {journal}
  {\bibinfo  {journal} {Phys. Rev. Lett.}\ }\textbf {\bibinfo {volume} {104}},\
  \bibinfo {pages} {137203} (\bibinfo {year} {2010}{\natexlab{a}})}\BibitemShut
  {NoStop}%
\bibitem [{\citenamefont {Udvardi}\ and\ \citenamefont
  {Szunyogh}(2009)}]{Udvardi2009}%
  \BibitemOpen
  \bibfield  {author} {\bibinfo {author} {\bibfnamefont {L.}~\bibnamefont
  {Udvardi}}\ and\ \bibinfo {author} {\bibfnamefont {L.}~\bibnamefont
  {Szunyogh}},\ }\bibfield  {title} {\bibinfo {title} {Chiral asymmetry of the
  spin-wave spectra in ultrathin magnetic films},\ }\href
  {https://doi.org/10.1103/PhysRevLett.102.207204} {\bibfield  {journal}
  {\bibinfo  {journal} {Phys. Rev. Lett.}\ }\textbf {\bibinfo {volume} {102}},\
  \bibinfo {pages} {207204} (\bibinfo {year} {2009})}\BibitemShut {NoStop}%
\bibitem [{\citenamefont {Costa}\ \emph {et~al.}(2010)\citenamefont {Costa},
  \citenamefont {Muniz}, \citenamefont {Lounis}, \citenamefont {Klautau},\ and\
  \citenamefont {Mills}}]{Costa2010a}%
  \BibitemOpen
  \bibfield  {author} {\bibinfo {author} {\bibfnamefont {A.~T.}\ \bibnamefont
  {Costa}}, \bibinfo {author} {\bibfnamefont {R.~B.}\ \bibnamefont {Muniz}},
  \bibinfo {author} {\bibfnamefont {S.}~\bibnamefont {Lounis}}, \bibinfo
  {author} {\bibfnamefont {A.~B.}\ \bibnamefont {Klautau}},\ and\ \bibinfo
  {author} {\bibfnamefont {D.~L.}\ \bibnamefont {Mills}},\ }\bibfield  {title}
  {\bibinfo {title} {Spin-orbit coupling and spin waves in ultrathin
  ferromagnets: {T}he spin-wave {R}ashba effect},\ }\href@noop {} {\bibfield
  {journal} {\bibinfo  {journal} {Phys. Rev. B}\ }\textbf {\bibinfo {volume}
  {82}},\ \bibinfo {pages} {014428} (\bibinfo {year} {2010})}\BibitemShut
  {NoStop}%
\bibitem [{\citenamefont {Costa}\ \emph {et~al.}(2020)\citenamefont {Costa},
  \citenamefont {Peres}, \citenamefont {Fernandez-Rossier},\ and\ \citenamefont
  {Costa}}]{Costa2020}%
  \BibitemOpen
  \bibfield  {author} {\bibinfo {author} {\bibfnamefont {M.}~\bibnamefont
  {Costa}}, \bibinfo {author} {\bibfnamefont {N.~M.~R.}\ \bibnamefont {Peres}},
  \bibinfo {author} {\bibfnamefont {J.}~\bibnamefont {Fernandez-Rossier}},\
  and\ \bibinfo {author} {\bibfnamefont {A.~T.}\ \bibnamefont {Costa}},\
  }\bibfield  {title} {\bibinfo {title} {Nonreciprocal magnons in a
  two-dimensional crystal with out-of-plane magnetization},\ }\href
  {https://doi.org/10.1103/physrevb.102.014450} {\bibfield  {journal} {\bibinfo
   {journal} {Physical Review B}\ }\textbf {\bibinfo {volume} {102}},\ \bibinfo
  {pages} {014450} (\bibinfo {year} {2020})}\BibitemShut {NoStop}%
\bibitem [{\citenamefont {Buczek}\ \emph {et~al.}(2011)\citenamefont {Buczek},
  \citenamefont {Ernst},\ and\ \citenamefont {Sandratskii}}]{Buczek2011}%
  \BibitemOpen
  \bibfield  {author} {\bibinfo {author} {\bibfnamefont {P.}~\bibnamefont
  {Buczek}}, \bibinfo {author} {\bibfnamefont {A.}~\bibnamefont {Ernst}},\ and\
  \bibinfo {author} {\bibfnamefont {L.~M.}\ \bibnamefont {Sandratskii}},\
  }\bibfield  {title} {\bibinfo {title} {Interface electronic complexes and
  landau damping of magnons in ultrathin magnets},\ }\href
  {https://doi.org/10.1103/PhysRevLett.106.157204} {\bibfield  {journal}
  {\bibinfo  {journal} {Phys. Rev. Lett.}\ }\textbf {\bibinfo {volume} {106}},\
  \bibinfo {pages} {157204} (\bibinfo {year} {2011})}\BibitemShut {NoStop}%
\bibitem [{\citenamefont {Qin}\ \emph {et~al.}(2015)\citenamefont {Qin},
  \citenamefont {Zakeri}, \citenamefont {Ernst}, \citenamefont {Sandratskii},
  \citenamefont {Buczek}, \citenamefont {Marmodoro}, \citenamefont {Chuang},
  \citenamefont {Zhang},\ and\ \citenamefont {Kirschner}}]{Qin2015}%
  \BibitemOpen
  \bibfield  {author} {\bibinfo {author} {\bibfnamefont {H.~J.}\ \bibnamefont
  {Qin}}, \bibinfo {author} {\bibfnamefont {K.}~\bibnamefont {Zakeri}},
  \bibinfo {author} {\bibfnamefont {A.}~\bibnamefont {Ernst}}, \bibinfo
  {author} {\bibfnamefont {L.~M.}\ \bibnamefont {Sandratskii}}, \bibinfo
  {author} {\bibfnamefont {P.}~\bibnamefont {Buczek}}, \bibinfo {author}
  {\bibfnamefont {A.}~\bibnamefont {Marmodoro}}, \bibinfo {author}
  {\bibfnamefont {T.~H.}\ \bibnamefont {Chuang}}, \bibinfo {author}
  {\bibfnamefont {Y.}~\bibnamefont {Zhang}},\ and\ \bibinfo {author}
  {\bibfnamefont {J.}~\bibnamefont {Kirschner}},\ }\bibfield  {title} {\bibinfo
  {title} {Long-living terahertz magnons in ultrathin metallic ferromagnets},\
  }\href {https://doi.org/doi.org/10.1038/ncomms7126} {\bibfield  {journal}
  {\bibinfo  {journal} {Nat. Commun.}\ }\textbf {\bibinfo {volume} {6}},\
  \bibinfo {pages} {6126} (\bibinfo {year} {2015})}\BibitemShut {NoStop}%
\bibitem [{\citenamefont {Zakeri}\ \emph
  {et~al.}(2021{\natexlab{b}})\citenamefont {Zakeri}, \citenamefont {Hjelt},
  \citenamefont {Maznichenko}, \citenamefont {Buczek},\ and\ \citenamefont
  {Ernst}}]{Zakeri2021a}%
  \BibitemOpen
  \bibfield  {author} {\bibinfo {author} {\bibfnamefont {K.}~\bibnamefont
  {Zakeri}}, \bibinfo {author} {\bibfnamefont {A.}~\bibnamefont {Hjelt}},
  \bibinfo {author} {\bibfnamefont {I.}~\bibnamefont {Maznichenko}}, \bibinfo
  {author} {\bibfnamefont {P.}~\bibnamefont {Buczek}},\ and\ \bibinfo {author}
  {\bibfnamefont {A.}~\bibnamefont {Ernst}},\ }\bibfield  {title} {\bibinfo
  {title} {Nonlinear decay of quantum confined magnons in itinerant
  ferromagnets},\ }\href {https://doi.org/10.1103/physrevlett.126.177203}
  {\bibfield  {journal} {\bibinfo  {journal} {Phys. Rev. Lett.}\ }\textbf
  {\bibinfo {volume} {126}},\ \bibinfo {pages} {177203} (\bibinfo {year}
  {2021}{\natexlab{b}})}\BibitemShut {NoStop}%
\bibitem [{\citenamefont {Paischer}\ \emph {et~al.}(2024)\citenamefont
  {Paischer}, \citenamefont {Eilmsteiner}, \citenamefont {Maznichenko},
  \citenamefont {Buczek}, \citenamefont {Zakeri}, \citenamefont {Ernst},\ and\
  \citenamefont {Buczek}}]{Paischer2024}%
  \BibitemOpen
  \bibfield  {author} {\bibinfo {author} {\bibfnamefont {S.}~\bibnamefont
  {Paischer}}, \bibinfo {author} {\bibfnamefont {D.}~\bibnamefont
  {Eilmsteiner}}, \bibinfo {author} {\bibfnamefont {I.}~\bibnamefont
  {Maznichenko}}, \bibinfo {author} {\bibfnamefont {N.}~\bibnamefont {Buczek}},
  \bibinfo {author} {\bibfnamefont {K.}~\bibnamefont {Zakeri}}, \bibinfo
  {author} {\bibfnamefont {A.}~\bibnamefont {Ernst}},\ and\ \bibinfo {author}
  {\bibfnamefont {P.~A.}\ \bibnamefont {Buczek}},\ }\bibfield  {title}
  {\bibinfo {title} {Correlations, disorder, and multimagnon processes in
  terahertz spin dynamics of magnetic nanostructures: {A} first-principles
  investigation},\ }\href {https://doi.org/10.1103/physrevb.109.l220405}
  {\bibfield  {journal} {\bibinfo  {journal} {Physical Review B}\ }\textbf
  {\bibinfo {volume} {109}},\ \bibinfo {pages} {l220405} (\bibinfo {year}
  {2024})}\BibitemShut {NoStop}%
\bibitem [{\citenamefont {Zakeri}\ \emph {et~al.}(2012)\citenamefont {Zakeri},
  \citenamefont {Zhang}, \citenamefont {Chuang},\ and\ \citenamefont
  {Kirschner}}]{Zakeri2012}%
  \BibitemOpen
  \bibfield  {author} {\bibinfo {author} {\bibfnamefont {K.}~\bibnamefont
  {Zakeri}}, \bibinfo {author} {\bibfnamefont {Y.}~\bibnamefont {Zhang}},
  \bibinfo {author} {\bibfnamefont {T.-H.}\ \bibnamefont {Chuang}},\ and\
  \bibinfo {author} {\bibfnamefont {J.}~\bibnamefont {Kirschner}},\ }\bibfield
  {title} {\bibinfo {title} {Magnon lifetimes on the {F}e(110) surface: The
  role of spin-orbit coupling},\ }\href
  {https://doi.org/10.1103/PhysRevLett.108.197205} {\bibfield  {journal}
  {\bibinfo  {journal} {Phys. Rev. Lett.}\ }\textbf {\bibinfo {volume} {108}},\
  \bibinfo {pages} {197205} (\bibinfo {year} {2012})}\BibitemShut {NoStop}%
\bibitem [{\citenamefont {Zakeri}\ and\ \citenamefont {von
  Faber}(2024)}]{Zakeri2024}%
  \BibitemOpen
  \bibfield  {author} {\bibinfo {author} {\bibfnamefont {K.}~\bibnamefont
  {Zakeri}}\ and\ \bibinfo {author} {\bibfnamefont {A.}~\bibnamefont {von
  Faber}},\ }\bibfield  {title} {\bibinfo {title} {Giant spin-orbit induced
  magnon nonreciprocity in ultrathin ferromagnets},\ }\href
  {https://doi.org/10.1103/physrevlett.132.126702} {\bibfield  {journal}
  {\bibinfo  {journal} {Phys. Rev. Lett.}\ }\textbf {\bibinfo {volume} {132}},\
  \bibinfo {pages} {126702} (\bibinfo {year} {2024})}\BibitemShut {NoStop}%
\bibitem [{\citenamefont {Zakeri}(2020)}]{Zakeri2020}%
  \BibitemOpen
  \bibfield  {author} {\bibinfo {author} {\bibfnamefont {K.}~\bibnamefont
  {Zakeri}},\ }\bibfield  {title} {\bibinfo {title} {Magnonic crystals:
  {T}owards terahertz frequencies},\ }\href
  {https://doi.org/10.1088/1361-648x/ab88f2} {\bibfield  {journal} {\bibinfo
  {journal} {Journal of Physics: Condensed Matter}\ }\textbf {\bibinfo {volume}
  {32}},\ \bibinfo {pages} {363001} (\bibinfo {year} {2020})}\BibitemShut
  {NoStop}%
\bibitem [{\citenamefont {Zakeri}(2018)}]{Zakeri2018}%
  \BibitemOpen
  \bibfield  {author} {\bibinfo {author} {\bibfnamefont {K.}~\bibnamefont
  {Zakeri}},\ }\bibfield  {title} {\bibinfo {title} {Terahertz magnonics:
  Feasibility of using terahertz magnons for information processing},\ }\href
  {https://doi.org/10.1016/j.physc.2018.02.035} {\bibfield  {journal} {\bibinfo
   {journal} {Physica C-superconductivity and Its Applications}\ }\textbf
  {\bibinfo {volume} {549}},\ \bibinfo {pages} {164} (\bibinfo {year}
  {2018})}\BibitemShut {NoStop}%
\bibitem [{\citenamefont {Wang}\ \emph {et~al.}(1979)\citenamefont {Wang},
  \citenamefont {Dunlap}, \citenamefont {Celotta},\ and\ \citenamefont
  {Pierce}}]{Wang1979}%
  \BibitemOpen
  \bibfield  {author} {\bibinfo {author} {\bibfnamefont {G.~C.}\ \bibnamefont
  {Wang}}, \bibinfo {author} {\bibfnamefont {B.~I.}\ \bibnamefont {Dunlap}},
  \bibinfo {author} {\bibfnamefont {R.~J.}\ \bibnamefont {Celotta}},\ and\
  \bibinfo {author} {\bibfnamefont {D.~T.}\ \bibnamefont {Pierce}},\ }\bibfield
   {title} {\bibinfo {title} {Symmetry in low-energy-polarized-electron
  diffraction},\ }\href {https://doi.org/10.1103/PhysRevLett.42.1349}
  {\bibfield  {journal} {\bibinfo  {journal} {Phys. Rev. Lett.}\ }\textbf
  {\bibinfo {volume} {42}},\ \bibinfo {pages} {1349} (\bibinfo {year}
  {1979})}\BibitemShut {NoStop}%
\bibitem [{\citenamefont {Kirschner}(1985{\natexlab{a}})}]{Kirschner1985}%
  \BibitemOpen
  \bibfield  {author} {\bibinfo {author} {\bibfnamefont {J.}~\bibnamefont
  {Kirschner}},\ }\href {https://doi.org/10.1007/BFb0108668} {\emph {\bibinfo
  {title} {Polarized Electrons at Surfaces}}},\ \bibinfo {edition} {1st}\ ed.,\
  \bibinfo {series} {Springer Tracts in Modern Physics}, Vol.\ \bibinfo
  {volume} {106}\ (\bibinfo  {publisher} {Springer},\ \bibinfo {address}
  {Berlin Heidelberg},\ \bibinfo {year} {1985})\ pp.\ \bibinfo {pages}
  {1--160}\BibitemShut {NoStop}%
\bibitem [{\citenamefont {Feder}(1986)}]{Feder1986}%
  \BibitemOpen
  \bibfield  {author} {\bibinfo {author} {\bibfnamefont {R.}~\bibnamefont
  {Feder}},\ }\bibfield  {title} {\bibinfo {title} {Principles and theory of
  electron scattering and photoemission},\ }in\ \href
  {https://doi.org/10.1142/9789814415262_0004} {\emph {\bibinfo {booktitle}
  {Polarized Electrons in Surface Physics}}}\ (\bibinfo  {publisher} {{WORLD}
  {SCIENTIFIC}},\ \bibinfo {year} {1986})\ pp.\ \bibinfo {pages}
  {125--241}\BibitemShut {NoStop}%
\bibitem [{\citenamefont {Zakeri}\ and\ \citenamefont
  {Berthod}(2022)}]{Zakeri2022}%
  \BibitemOpen
  \bibfield  {author} {\bibinfo {author} {\bibfnamefont {K.}~\bibnamefont
  {Zakeri}}\ and\ \bibinfo {author} {\bibfnamefont {C.}~\bibnamefont
  {Berthod}},\ }\bibfield  {title} {\bibinfo {title} {Theory of spin-polarized
  high-resolution electron energy loss spectroscopy from nonmagnetic surfaces
  with a large spin-orbit coupling},\ }\href
  {https://doi.org/10.1103/physrevb.106.235117} {\bibfield  {journal} {\bibinfo
   {journal} {Physical Review B}\ }\textbf {\bibinfo {volume} {106}},\ \bibinfo
  {pages} {235117} (\bibinfo {year} {2022})}\BibitemShut {NoStop}%
\bibitem [{\citenamefont {Elmers}(2007)}]{Elmers2007}%
  \BibitemOpen
  \bibfield  {author} {\bibinfo {author} {\bibfnamefont {H.-J.}\ \bibnamefont
  {Elmers}},\ }\href {https://doi.org/10.1002/9780470022184.hmm307} {\bibinfo
  {title} {Spin-polarized low energy electron diffraction}} (\bibinfo {year}
  {2007})\BibitemShut {NoStop}%
\bibitem [{\citenamefont {Zakeri}(2017)}]{Zakeri2017}%
  \BibitemOpen
  \bibfield  {author} {\bibinfo {author} {\bibfnamefont {K.}~\bibnamefont
  {Zakeri}},\ }\bibfield  {title} {\bibinfo {title} {Probing of the interfacial
  {H}eisenberg and {D}zyaloshinskii--{M}oriya exchange interaction by magnon
  spectroscopy},\ }\href {https://doi.org/10.1088/0953-8984/29/1/013001}
  {\bibfield  {journal} {\bibinfo  {journal} {Journal of Physics: Condensed
  Matter}\ }\textbf {\bibinfo {volume} {29}},\ \bibinfo {pages} {013001}
  (\bibinfo {year} {2017})}\BibitemShut {NoStop}%
\bibitem [{\citenamefont {Zakeri}\ \emph
  {et~al.}(2024{\natexlab{a}})\citenamefont {Zakeri}, \citenamefont {von
  Faber}, \citenamefont {Mankovsky},\ and\ \citenamefont
  {Ebert}}]{Zakeri2024b}%
  \BibitemOpen
  \bibfield  {author} {\bibinfo {author} {\bibfnamefont {K.}~\bibnamefont
  {Zakeri}}, \bibinfo {author} {\bibfnamefont {A.}~\bibnamefont {von Faber}},
  \bibinfo {author} {\bibfnamefont {S.}~\bibnamefont {Mankovsky}},\ and\
  \bibinfo {author} {\bibfnamefont {H.}~\bibnamefont {Ebert}},\ }\bibfield
  {title} {\bibinfo {title} {Unraveling the complexity of the
  dzyaloshinskii-moriya interaction in layered magnets: Towards its full
  magnitude and chirality control},\ }\bibfield  {journal} {\bibinfo  {journal}
  {arXiv.2402.18466}\ }\href {https://doi.org/10.48550/arXiv.2402.18466}
  {10.48550/arXiv.2402.18466} (\bibinfo {year}
  {2024}{\natexlab{a}})\BibitemShut {NoStop}%
\bibitem [{Note1()}]{Note1}%
  \BibitemOpen
  \bibinfo {note} {The term double layer refers to the case in which the sample
  is composed of two atomic layers of the same material. Likewise, the term
  bilayer refers to the case in which the sample is composed of two atomic
  layers made of two different materials.}\BibitemShut {Stop}%
\bibitem [{\citenamefont {Heinz}\ and\ \citenamefont
  {Hammer}(2009)}]{Heinz2009}%
  \BibitemOpen
  \bibfield  {author} {\bibinfo {author} {\bibfnamefont {K.}~\bibnamefont
  {Heinz}}\ and\ \bibinfo {author} {\bibfnamefont {L.}~\bibnamefont {Hammer}},\
  }\bibfield  {title} {\bibinfo {title} {Nanostructure formation on
  {I}r(100)},\ }\href {https://doi.org/10.1016/j.progsurf.2008.10.003}
  {\bibfield  {journal} {\bibinfo  {journal} {Prog. Surf. Sci.}\ }\textbf
  {\bibinfo {volume} {84}},\ \bibinfo {pages} {2 } (\bibinfo {year}
  {2009})}\BibitemShut {NoStop}%
\bibitem [{\citenamefont {Zakeri}\ \emph
  {et~al.}(2021{\natexlab{c}})\citenamefont {Zakeri}, \citenamefont {Qin},\
  and\ \citenamefont {Ernst}}]{Zakeri2021}%
  \BibitemOpen
  \bibfield  {author} {\bibinfo {author} {\bibfnamefont {K.}~\bibnamefont
  {Zakeri}}, \bibinfo {author} {\bibfnamefont {H.}~\bibnamefont {Qin}},\ and\
  \bibinfo {author} {\bibfnamefont {A.}~\bibnamefont {Ernst}},\ }\bibfield
  {title} {\bibinfo {title} {Unconventional magnonic surface and interface
  states in layered ferromagnets},\ }\href
  {https://doi.org/10.1038/s42005-021-00521-7} {\bibfield  {journal} {\bibinfo
  {journal} {Commun. Phys.}\ }\textbf {\bibinfo {volume} {4}},\ \bibinfo
  {pages} {18} (\bibinfo {year} {2021}{\natexlab{c}})}\BibitemShut {NoStop}%
\bibitem [{\citenamefont {Zakeri}\ \emph
  {et~al.}(2023{\natexlab{a}})\citenamefont {Zakeri}, \citenamefont
  {Marmodoro}, \citenamefont {von Faber}, \citenamefont {Mankovsky},\ and\
  \citenamefont {Ebert}}]{Zakeri2023b}%
  \BibitemOpen
  \bibfield  {author} {\bibinfo {author} {\bibfnamefont {K.}~\bibnamefont
  {Zakeri}}, \bibinfo {author} {\bibfnamefont {A.}~\bibnamefont {Marmodoro}},
  \bibinfo {author} {\bibfnamefont {A.}~\bibnamefont {von Faber}}, \bibinfo
  {author} {\bibfnamefont {S.}~\bibnamefont {Mankovsky}},\ and\ \bibinfo
  {author} {\bibfnamefont {H.}~\bibnamefont {Ebert}},\ }\bibfield  {title}
  {\bibinfo {title} {Chirality-inverted {D}zyaloshinskii-{M}oriya
  interaction},\ }\href {https://doi.org/10.1103/physrevb.108.l100403}
  {\bibfield  {journal} {\bibinfo  {journal} {Physical Review B}\ }\textbf
  {\bibinfo {volume} {108}},\ \bibinfo {pages} {l100403} (\bibinfo {year}
  {2023}{\natexlab{a}})}\BibitemShut {NoStop}%
\bibitem [{\citenamefont {Zakeri}\ \emph
  {et~al.}(2024{\natexlab{b}})\citenamefont {Zakeri}, \citenamefont {von
  Faber},\ and\ \citenamefont {Ernst}}]{Zakeri2024a}%
  \BibitemOpen
  \bibfield  {author} {\bibinfo {author} {\bibfnamefont {K.}~\bibnamefont
  {Zakeri}}, \bibinfo {author} {\bibfnamefont {A.}~\bibnamefont {von Faber}},\
  and\ \bibinfo {author} {\bibfnamefont {A.}~\bibnamefont {Ernst}},\ }\bibfield
   {title} {\bibinfo {title} {Magnons and fundamental magnetic interactions in
  a ferromagnetic monolayer: The case of {N}i monolayer},\ }\bibfield
  {journal} {\bibinfo  {journal} {arXiv.2402.15251}\ }\href
  {https://doi.org/10.48550/arXiv.2402.15251} {10.48550/arXiv.2402.15251}
  (\bibinfo {year} {2024}{\natexlab{b}})\BibitemShut {NoStop}%
\bibitem [{\citenamefont {Tian}\ \emph {et~al.}(2009)\citenamefont {Tian},
  \citenamefont {Sander},\ and\ \citenamefont {Kirschner}}]{Tian2009}%
  \BibitemOpen
  \bibfield  {author} {\bibinfo {author} {\bibfnamefont {Z.}~\bibnamefont
  {Tian}}, \bibinfo {author} {\bibfnamefont {D.}~\bibnamefont {Sander}},\ and\
  \bibinfo {author} {\bibfnamefont {J.}~\bibnamefont {Kirschner}},\ }\bibfield
  {title} {\bibinfo {title} {Nonlinear magnetoelastic coupling of epitaxial
  layers of {F}e, {C}o, and {N}i on {I}r(100)},\ }\href
  {https://doi.org/10.1103/physrevb.79.024432} {\bibfield  {journal} {\bibinfo
  {journal} {Physical Review B}\ }\textbf {\bibinfo {volume} {79}},\ \bibinfo
  {pages} {024432} (\bibinfo {year} {2009})}\BibitemShut {NoStop}%
\bibitem [{\citenamefont {Qin}\ \emph {et~al.}(2013)\citenamefont {Qin},
  \citenamefont {Zakeri}, \citenamefont {Ernst}, \citenamefont {Chuang},
  \citenamefont {Chen}, \citenamefont {Meng},\ and\ \citenamefont
  {Kirschner}}]{Qin2013}%
  \BibitemOpen
  \bibfield  {author} {\bibinfo {author} {\bibfnamefont {H.~J.}\ \bibnamefont
  {Qin}}, \bibinfo {author} {\bibfnamefont {K.}~\bibnamefont {Zakeri}},
  \bibinfo {author} {\bibfnamefont {A.}~\bibnamefont {Ernst}}, \bibinfo
  {author} {\bibfnamefont {T.-H.}\ \bibnamefont {Chuang}}, \bibinfo {author}
  {\bibfnamefont {Y.-J.}\ \bibnamefont {Chen}}, \bibinfo {author}
  {\bibfnamefont {Y.}~\bibnamefont {Meng}},\ and\ \bibinfo {author}
  {\bibfnamefont {J.}~\bibnamefont {Kirschner}},\ }\bibfield  {title} {\bibinfo
  {title} {Magnons in ultrathin ferromagnetic films with a large perpendicular
  magnetic anisotropy},\ }\href {https://doi.org/10.1103/PhysRevB.88.020404}
  {\bibfield  {journal} {\bibinfo  {journal} {Phys. Rev. B}\ }\textbf {\bibinfo
  {volume} {88}},\ \bibinfo {pages} {020404} (\bibinfo {year}
  {2013})}\BibitemShut {NoStop}%
\bibitem [{\citenamefont {Pratzer}\ \emph {et~al.}(2003)\citenamefont
  {Pratzer}, \citenamefont {Elmers},\ and\ \citenamefont
  {Getzlaff}}]{Pratzer2003}%
  \BibitemOpen
  \bibfield  {author} {\bibinfo {author} {\bibfnamefont {M.}~\bibnamefont
  {Pratzer}}, \bibinfo {author} {\bibfnamefont {H.~J.}\ \bibnamefont
  {Elmers}},\ and\ \bibinfo {author} {\bibfnamefont {M.}~\bibnamefont
  {Getzlaff}},\ }\bibfield  {title} {\bibinfo {title} {Heteroepitaxial growth
  of {C}o on {W}(110) investigated by scanning tunneling microscopy},\ }\href
  {https://doi.org/10.1103/PhysRevB.67.153405} {\bibfield  {journal} {\bibinfo
  {journal} {Phys. Rev. B}\ }\textbf {\bibinfo {volume} {67}},\ \bibinfo
  {pages} {153405} (\bibinfo {year} {2003})}\BibitemShut {NoStop}%
\bibitem [{\citenamefont {Etzkorn}\ \emph {et~al.}(2005)\citenamefont
  {Etzkorn}, \citenamefont {Anil~Kumar}, \citenamefont {Tang}, \citenamefont
  {Zhang},\ and\ \citenamefont {Kirschner}}]{Etzkorn2005}%
  \BibitemOpen
  \bibfield  {author} {\bibinfo {author} {\bibfnamefont {M.}~\bibnamefont
  {Etzkorn}}, \bibinfo {author} {\bibfnamefont {P.~S.}\ \bibnamefont
  {Anil~Kumar}}, \bibinfo {author} {\bibfnamefont {W.}~\bibnamefont {Tang}},
  \bibinfo {author} {\bibfnamefont {Y.}~\bibnamefont {Zhang}},\ and\ \bibinfo
  {author} {\bibfnamefont {J.}~\bibnamefont {Kirschner}},\ }\bibfield  {title}
  {\bibinfo {title} {High-wave-vector spin waves in ultrathin {Co} films on
  {W}(110)},\ }\href {https://doi.org/10.1103/PhysRevB.72.184420} {\bibfield
  {journal} {\bibinfo  {journal} {Phys. Rev. B}\ }\textbf {\bibinfo {volume}
  {72}},\ \bibinfo {pages} {184420} (\bibinfo {year} {2005})}\BibitemShut
  {NoStop}%
\bibitem [{\citenamefont {Prokop}\ \emph {et~al.}(2009)\citenamefont {Prokop},
  \citenamefont {Tang}, \citenamefont {Zhang}, \citenamefont {Tudosa},
  \citenamefont {Peixoto}, \citenamefont {Zakeri},\ and\ \citenamefont
  {Kirschner}}]{Prokop2009}%
  \BibitemOpen
  \bibfield  {author} {\bibinfo {author} {\bibfnamefont {J.}~\bibnamefont
  {Prokop}}, \bibinfo {author} {\bibfnamefont {W.~X.}\ \bibnamefont {Tang}},
  \bibinfo {author} {\bibfnamefont {Y.}~\bibnamefont {Zhang}}, \bibinfo
  {author} {\bibfnamefont {I.}~\bibnamefont {Tudosa}}, \bibinfo {author}
  {\bibfnamefont {T.~R.~F.}\ \bibnamefont {Peixoto}}, \bibinfo {author}
  {\bibfnamefont {K.}~\bibnamefont {Zakeri}},\ and\ \bibinfo {author}
  {\bibfnamefont {J.}~\bibnamefont {Kirschner}},\ }\bibfield  {title} {\bibinfo
  {title} {Magnons in a ferromagnetic monolayer},\ }\href
  {https://doi.org/10.1103/PhysRevLett.102.177206} {\bibfield  {journal}
  {\bibinfo  {journal} {Phys. Rev. Lett.}\ }\textbf {\bibinfo {volume} {102}},\
  \bibinfo {pages} {177206} (\bibinfo {year} {2009})}\BibitemShut {NoStop}%
\bibitem [{\citenamefont {Zakeri}\ \emph {et~al.}(2011)\citenamefont {Zakeri},
  \citenamefont {Zhang}, \citenamefont {Prokop}, \citenamefont {Chuang},
  \citenamefont {Tang},\ and\ \citenamefont {Kirschner}}]{Zakeri2011}%
  \BibitemOpen
  \bibfield  {author} {\bibinfo {author} {\bibfnamefont {K.}~\bibnamefont
  {Zakeri}}, \bibinfo {author} {\bibfnamefont {Y.}~\bibnamefont {Zhang}},
  \bibinfo {author} {\bibfnamefont {J.}~\bibnamefont {Prokop}}, \bibinfo
  {author} {\bibfnamefont {T.~H.}\ \bibnamefont {Chuang}}, \bibinfo {author}
  {\bibfnamefont {W.~X.}\ \bibnamefont {Tang}},\ and\ \bibinfo {author}
  {\bibfnamefont {J.}~\bibnamefont {Kirschner}},\ }\bibfield  {title} {\bibinfo
  {title} {Magnon excitations in ultrathin {F}e layers: The influence of the
  {D}zyaloshinskii-{M}oriya interaction},\ }\href
  {https://doi.org/10.1088/1742-6596/303/1/012004} {\bibfield  {journal}
  {\bibinfo  {journal} {Journal of Physics: Conference Series}\ }\textbf
  {\bibinfo {volume} {303}},\ \bibinfo {pages} {012004} (\bibinfo {year}
  {2011})}\BibitemShut {NoStop}%
\bibitem [{\citenamefont {Tsurkan}\ and\ \citenamefont
  {Zakeri}(2020)}]{Tsurkan2020}%
  \BibitemOpen
  \bibfield  {author} {\bibinfo {author} {\bibfnamefont {S.}~\bibnamefont
  {Tsurkan}}\ and\ \bibinfo {author} {\bibfnamefont {K.}~\bibnamefont
  {Zakeri}},\ }\bibfield  {title} {\bibinfo {title} {Giant
  {Dzyaloshinskii-Moriya} interaction in epitaxial {Co/Fe} bilayers with
  {$C_{2v}$} symmetry},\ }\href {https://doi.org/10.1103/physrevb.102.060406}
  {\bibfield  {journal} {\bibinfo  {journal} {Physical Review B}\ }\textbf
  {\bibinfo {volume} {102}},\ \bibinfo {pages} {060406} (\bibinfo {year}
  {2020})}\BibitemShut {NoStop}%
\bibitem [{\citenamefont {Chuang}\ \emph {et~al.}(2014)\citenamefont {Chuang},
  \citenamefont {Zakeri}, \citenamefont {Ernst}, \citenamefont {Zhang},
  \citenamefont {Qin}, \citenamefont {Meng}, \citenamefont {Chen},\ and\
  \citenamefont {Kirschner}}]{Chuang2014}%
  \BibitemOpen
  \bibfield  {author} {\bibinfo {author} {\bibfnamefont {T.-H.}\ \bibnamefont
  {Chuang}}, \bibinfo {author} {\bibfnamefont {K.}~\bibnamefont {Zakeri}},
  \bibinfo {author} {\bibfnamefont {A.}~\bibnamefont {Ernst}}, \bibinfo
  {author} {\bibfnamefont {Y.}~\bibnamefont {Zhang}}, \bibinfo {author}
  {\bibfnamefont {H.~J.}\ \bibnamefont {Qin}}, \bibinfo {author} {\bibfnamefont
  {Y.}~\bibnamefont {Meng}}, \bibinfo {author} {\bibfnamefont {Y.-J.}\
  \bibnamefont {Chen}},\ and\ \bibinfo {author} {\bibfnamefont
  {J.}~\bibnamefont {Kirschner}},\ }\bibfield  {title} {\bibinfo {title}
  {Magnetic properties and magnon excitations in {Fe(001)} films grown on
  {Ir(001)}},\ }\href {https://doi.org/10.1103/PhysRevB.89.174404} {\bibfield
  {journal} {\bibinfo  {journal} {Phys. Rev. B}\ }\textbf {\bibinfo {volume}
  {89}},\ \bibinfo {pages} {174404} (\bibinfo {year} {2014})}\BibitemShut
  {NoStop}%
\bibitem [{\citenamefont {Zakeri}\ \emph
  {et~al.}(2010{\natexlab{b}})\citenamefont {Zakeri}, \citenamefont {Peixoto},
  \citenamefont {Zhang}, \citenamefont {Prokop},\ and\ \citenamefont
  {Kirschner}}]{Zakeri2010a}%
  \BibitemOpen
  \bibfield  {author} {\bibinfo {author} {\bibfnamefont {K.}~\bibnamefont
  {Zakeri}}, \bibinfo {author} {\bibfnamefont {T.}~\bibnamefont {Peixoto}},
  \bibinfo {author} {\bibfnamefont {Y.}~\bibnamefont {Zhang}}, \bibinfo
  {author} {\bibfnamefont {J.}~\bibnamefont {Prokop}},\ and\ \bibinfo {author}
  {\bibfnamefont {J.}~\bibnamefont {Kirschner}},\ }\bibfield  {title} {\bibinfo
  {title} {On the preparation of clean tungsten single crystals},\ }\href
  {https://doi.org/10.1016/j.susc.2009.10.020} {\bibfield  {journal} {\bibinfo
  {journal} {Surface Science}\ }\textbf {\bibinfo {volume} {604}},\ \bibinfo
  {pages} {L1} (\bibinfo {year} {2010}{\natexlab{b}})}\BibitemShut {NoStop}%
\bibitem [{\citenamefont {Vollmer}\ \emph {et~al.}(2003)\citenamefont
  {Vollmer}, \citenamefont {Etzkorn}, \citenamefont {Kumar}, \citenamefont
  {Ibach},\ and\ \citenamefont {Kirschner}}]{Vollmer2003}%
  \BibitemOpen
  \bibfield  {author} {\bibinfo {author} {\bibfnamefont {R.}~\bibnamefont
  {Vollmer}}, \bibinfo {author} {\bibfnamefont {M.}~\bibnamefont {Etzkorn}},
  \bibinfo {author} {\bibfnamefont {P.~S.~A.}\ \bibnamefont {Kumar}}, \bibinfo
  {author} {\bibfnamefont {H.}~\bibnamefont {Ibach}},\ and\ \bibinfo {author}
  {\bibfnamefont {J.}~\bibnamefont {Kirschner}},\ }\bibfield  {title} {\bibinfo
  {title} {Spin-polarized electron energy loss spectroscopy of high energy,
  large wave vector spin waves in ultrathin fcc {Co} films on {Cu(001)}},\
  }\href {https://doi.org/10.1103/PhysRevLett.91.147201} {\bibfield  {journal}
  {\bibinfo  {journal} {Phys. Rev. Lett.}\ }\textbf {\bibinfo {volume} {91}},\
  \bibinfo {pages} {147201} (\bibinfo {year} {2003})}\BibitemShut {NoStop}%
\bibitem [{\citenamefont {Zakeri}\ and\ \citenamefont
  {Kirschner}(2013)}]{Zakeri2013}%
  \BibitemOpen
  \bibfield  {author} {\bibinfo {author} {\bibfnamefont {K.}~\bibnamefont
  {Zakeri}}\ and\ \bibinfo {author} {\bibfnamefont {J.}~\bibnamefont
  {Kirschner}},\ }\bibinfo {title} {Probing magnons by spin-polarized
  electrons}\ (\bibinfo  {publisher} {Springer},\ \bibinfo {address} {Berlin,
  Heidelberg},\ \bibinfo {year} {2013})\ Chap.~\bibinfo {chapter} {7}, pp.\
  \bibinfo {pages} {84 -- 99}\BibitemShut {NoStop}%
\bibitem [{\citenamefont {Zakeri}\ \emph {et~al.}(2013)\citenamefont {Zakeri},
  \citenamefont {Zhang},\ and\ \citenamefont {Kirschner}}]{Zakeri2013b}%
  \BibitemOpen
  \bibfield  {author} {\bibinfo {author} {\bibfnamefont {K.}~\bibnamefont
  {Zakeri}}, \bibinfo {author} {\bibfnamefont {Y.}~\bibnamefont {Zhang}},\ and\
  \bibinfo {author} {\bibfnamefont {J.}~\bibnamefont {Kirschner}},\ }\bibfield
  {title} {\bibinfo {title} {Surface magnons probed by spin-polarized electron
  energy loss spectroscopy},\ }\href
  {https://doi.org/10.1016/j.elspec.2012.06.009} {\bibfield  {journal}
  {\bibinfo  {journal} {Journal of Electron Spectroscopy and Related
  Phenomena}\ }\textbf {\bibinfo {volume} {189}},\ \bibinfo {pages} {157}
  (\bibinfo {year} {2013})}\BibitemShut {NoStop}%
\bibitem [{\citenamefont {Zhang}\ \emph {et~al.}(2011)\citenamefont {Zhang},
  \citenamefont {Ignatiev}, \citenamefont {Prokop}, \citenamefont {Tudosa},
  \citenamefont {Peixoto}, \citenamefont {Tang}, \citenamefont {Zakeri},
  \citenamefont {Stepanyuk},\ and\ \citenamefont {Kirschner}}]{Zhang2011}%
  \BibitemOpen
  \bibfield  {author} {\bibinfo {author} {\bibfnamefont {Y.}~\bibnamefont
  {Zhang}}, \bibinfo {author} {\bibfnamefont {P.~A.}\ \bibnamefont {Ignatiev}},
  \bibinfo {author} {\bibfnamefont {J.}~\bibnamefont {Prokop}}, \bibinfo
  {author} {\bibfnamefont {I.}~\bibnamefont {Tudosa}}, \bibinfo {author}
  {\bibfnamefont {T.~R.~F.}\ \bibnamefont {Peixoto}}, \bibinfo {author}
  {\bibfnamefont {W.~X.}\ \bibnamefont {Tang}}, \bibinfo {author}
  {\bibfnamefont {K.}~\bibnamefont {Zakeri}}, \bibinfo {author} {\bibfnamefont
  {V.~S.}\ \bibnamefont {Stepanyuk}},\ and\ \bibinfo {author} {\bibfnamefont
  {J.}~\bibnamefont {Kirschner}},\ }\bibfield  {title} {\bibinfo {title}
  {Elementary excitations at magnetic surfaces and their spin dependence.},\
  }\href {https://doi.org/10.1103/PhysRevLett.106.127201} {\bibfield  {journal}
  {\bibinfo  {journal} {Phys. Rev. Lett.}\ }\textbf {\bibinfo {volume} {106}},\
  \bibinfo {pages} {127201} (\bibinfo {year} {2011})}\BibitemShut {NoStop}%
\bibitem [{\citenamefont {Zakeri}\ \emph
  {et~al.}(2023{\natexlab{b}})\citenamefont {Zakeri}, \citenamefont {Rau},
  \citenamefont {Jandke}, \citenamefont {Yang}, \citenamefont {Wulfhekel},\
  and\ \citenamefont {Berthod}}]{Zakeri2023}%
  \BibitemOpen
  \bibfield  {author} {\bibinfo {author} {\bibfnamefont {K.}~\bibnamefont
  {Zakeri}}, \bibinfo {author} {\bibfnamefont {D.}~\bibnamefont {Rau}},
  \bibinfo {author} {\bibfnamefont {J.}~\bibnamefont {Jandke}}, \bibinfo
  {author} {\bibfnamefont {F.}~\bibnamefont {Yang}}, \bibinfo {author}
  {\bibfnamefont {W.}~\bibnamefont {Wulfhekel}},\ and\ \bibinfo {author}
  {\bibfnamefont {C.}~\bibnamefont {Berthod}},\ }\bibfield  {title} {\bibinfo
  {title} {Direct probing of a large spin-orbit coupling in the {FeSe}
  superconducting monolayer on {STO}},\ }\href
  {https://doi.org/10.1021/acsnano.3c02876} {\bibfield  {journal} {\bibinfo
  {journal} {ACS Nano}\ }\textbf {\bibinfo {volume} {17}},\ \bibinfo {pages}
  {9575} (\bibinfo {year} {2023}{\natexlab{b}})}\BibitemShut {NoStop}%
\bibitem [{\citenamefont {Gokhale}\ \emph {et~al.}(1992)\citenamefont
  {Gokhale}, \citenamefont {Ormeci},\ and\ \citenamefont
  {Mills}}]{Gokhale1992}%
  \BibitemOpen
  \bibfield  {author} {\bibinfo {author} {\bibfnamefont {M.~P.}\ \bibnamefont
  {Gokhale}}, \bibinfo {author} {\bibfnamefont {A.}~\bibnamefont {Ormeci}},\
  and\ \bibinfo {author} {\bibfnamefont {D.~L.}\ \bibnamefont {Mills}},\
  }\bibfield  {title} {\bibinfo {title} {Inelastic scattering of low-energy
  electrons by spin excitations on ferromagnets},\ }\href
  {https://doi.org/10.1103/physrevb.46.8978} {\bibfield  {journal} {\bibinfo
  {journal} {Physical Review B}\ }\textbf {\bibinfo {volume} {46}},\ \bibinfo
  {pages} {8978} (\bibinfo {year} {1992})}\BibitemShut {NoStop}%
\bibitem [{\citenamefont {Gokhale}\ and\ \citenamefont
  {Mills}(1994)}]{Gokhale1994}%
  \BibitemOpen
  \bibfield  {author} {\bibinfo {author} {\bibfnamefont {M.~P.}\ \bibnamefont
  {Gokhale}}\ and\ \bibinfo {author} {\bibfnamefont {D.~L.}\ \bibnamefont
  {Mills}},\ }\bibfield  {title} {\bibinfo {title} {Spin excitations of a model
  itinerant ferromagnetic film: {S}pin waves, {S}toner excitations, and
  spin-polarized electron-energy-loss spectroscopy},\ }\href
  {https://doi.org/10.1103/physrevb.49.3880} {\bibfield  {journal} {\bibinfo
  {journal} {Physical Review B}\ }\textbf {\bibinfo {volume} {49}},\ \bibinfo
  {pages} {3880} (\bibinfo {year} {1994})}\BibitemShut {NoStop}%
\bibitem [{\citenamefont {Plihal}\ \emph {et~al.}(1999)\citenamefont {Plihal},
  \citenamefont {Mills},\ and\ \citenamefont {Kirschner}}]{Plihal1999}%
  \BibitemOpen
  \bibfield  {author} {\bibinfo {author} {\bibfnamefont {M.}~\bibnamefont
  {Plihal}}, \bibinfo {author} {\bibfnamefont {D.~L.}\ \bibnamefont {Mills}},\
  and\ \bibinfo {author} {\bibfnamefont {J.}~\bibnamefont {Kirschner}},\
  }\bibfield  {title} {\bibinfo {title} {Spin wave signature in the spin
  polarized electron energy loss spectrum of ultrathin fe films: {T}heory and
  experiment},\ }\href {https://doi.org/10.1103/physrevlett.82.2579} {\bibfield
   {journal} {\bibinfo  {journal} {Physical Review Letters}\ }\textbf {\bibinfo
  {volume} {82}},\ \bibinfo {pages} {2579} (\bibinfo {year}
  {1999})}\BibitemShut {NoStop}%
\bibitem [{\citenamefont {Zhang}\ \emph {et~al.}(2012)\citenamefont {Zhang},
  \citenamefont {Chuang}, \citenamefont {Zakeri},\ and\ \citenamefont
  {Kirschner}}]{Zhang2012}%
  \BibitemOpen
  \bibfield  {author} {\bibinfo {author} {\bibfnamefont {Y.}~\bibnamefont
  {Zhang}}, \bibinfo {author} {\bibfnamefont {T.-H.}\ \bibnamefont {Chuang}},
  \bibinfo {author} {\bibfnamefont {K.}~\bibnamefont {Zakeri}},\ and\ \bibinfo
  {author} {\bibfnamefont {J.}~\bibnamefont {Kirschner}},\ }\bibfield  {title}
  {\bibinfo {title} {Relaxation time of terahertz magnons excited at
  ferromagnetic surfaces},\ }\href
  {https://doi.org/10.1103/PhysRevLett.109.087203} {\bibfield  {journal}
  {\bibinfo  {journal} {Phys. Rev. Lett.}\ }\textbf {\bibinfo {volume} {109}},\
  \bibinfo {pages} {087203} (\bibinfo {year} {2012})}\BibitemShut {NoStop}%
\bibitem [{\citenamefont {Kirschner}(1985{\natexlab{b}})}]{Kirschner1985a}%
  \BibitemOpen
  \bibfield  {author} {\bibinfo {author} {\bibfnamefont {J.}~\bibnamefont
  {Kirschner}},\ }\bibfield  {title} {\bibinfo {title} {Direct and exchange
  contributions in inelastic scattering of spin-polarized electrons from
  iron},\ }\href {https://doi.org/10.1103/physrevlett.55.973} {\bibfield
  {journal} {\bibinfo  {journal} {Physical Review Letters}\ }\textbf {\bibinfo
  {volume} {55}},\ \bibinfo {pages} {973} (\bibinfo {year}
  {1985}{\natexlab{b}})}\BibitemShut {NoStop}%
\bibitem [{\citenamefont {Gradmann}\ and\ \citenamefont
  {Alvarado}(1986)}]{Gradmann1986}%
  \BibitemOpen
  \bibfield  {author} {\bibinfo {author} {\bibfnamefont {U.}~\bibnamefont
  {Gradmann}}\ and\ \bibinfo {author} {\bibfnamefont {S.~F.}\ \bibnamefont
  {Alvarado}},\ }\bibfield  {title} {\bibinfo {title} {Elastic spin-polarized
  low-energy electron scattering from magnetic surfaces},\ }in\ \href
  {https://doi.org/10.1142/9789814415262_0007} {\emph {\bibinfo {booktitle}
  {Polarized Electrons in Surface Physics}}}\ (\bibinfo  {publisher} {{WORLD}
  {SCIENTIFIC}},\ \bibinfo {year} {1986})\ pp.\ \bibinfo {pages}
  {321--352}\BibitemShut {NoStop}%
\bibitem [{\citenamefont {Vollmer}\ \emph {et~al.}(2004)\citenamefont
  {Vollmer}, \citenamefont {Etzkorn}, \citenamefont {Kumar}, \citenamefont
  {Ibach},\ and\ \citenamefont {Kirschner}}]{Vollmer2004}%
  \BibitemOpen
  \bibfield  {author} {\bibinfo {author} {\bibfnamefont {R.}~\bibnamefont
  {Vollmer}}, \bibinfo {author} {\bibfnamefont {M.}~\bibnamefont {Etzkorn}},
  \bibinfo {author} {\bibfnamefont {P.~S.~A.}\ \bibnamefont {Kumar}}, \bibinfo
  {author} {\bibfnamefont {H.}~\bibnamefont {Ibach}},\ and\ \bibinfo {author}
  {\bibfnamefont {J.}~\bibnamefont {Kirschner}},\ }\bibfield  {title} {\bibinfo
  {title} {Spin-wave excitation observed by spin-polarized electron energy loss
  spectroscopy: {A} new method for the investigation of surface- and thin-film
  spin waves on the atomic scale},\ }\href
  {https://doi.org/10.1016/j.tsf.2004.06.029} {\bibfield  {journal} {\bibinfo
  {journal} {Thin Solid Films}\ }\textbf {\bibinfo {volume} {464 - 465}},\
  \bibinfo {pages} {42} (\bibinfo {year} {2004})}\BibitemShut {NoStop}%
\bibitem [{\citenamefont {Etzkorn}\ \emph {et~al.}(2004)\citenamefont
  {Etzkorn}, \citenamefont {Anil~Kumar}, \citenamefont {Vollmer}, \citenamefont
  {Ibach},\ and\ \citenamefont {Kirschner}}]{Etzkorn2004}%
  \BibitemOpen
  \bibfield  {author} {\bibinfo {author} {\bibfnamefont {M.}~\bibnamefont
  {Etzkorn}}, \bibinfo {author} {\bibfnamefont {P.~S.}\ \bibnamefont
  {Anil~Kumar}}, \bibinfo {author} {\bibfnamefont {R.}~\bibnamefont {Vollmer}},
  \bibinfo {author} {\bibfnamefont {H.}~\bibnamefont {Ibach}},\ and\ \bibinfo
  {author} {\bibfnamefont {J.}~\bibnamefont {Kirschner}},\ }\bibfield  {title}
  {\bibinfo {title} {Spin waves in ultrathin {C}o-films measured by spin
  polarized electron energy loss spectroscopy},\ }\href
  {https://doi.org/10.1016/j.susc.2004.05.051} {\bibfield  {journal} {\bibinfo
  {journal} {Surface Science}\ }\textbf {\bibinfo {volume} {566-568,}},\
  \bibinfo {pages} {241 } (\bibinfo {year} {2004})}\BibitemShut {NoStop}%
\bibitem [{\citenamefont {Tsui}\ \emph {et~al.}(1971)\citenamefont {Tsui},
  \citenamefont {Dietz},\ and\ \citenamefont {Walker}}]{Tsui1971}%
  \BibitemOpen
  \bibfield  {author} {\bibinfo {author} {\bibfnamefont {D.~C.}\ \bibnamefont
  {Tsui}}, \bibinfo {author} {\bibfnamefont {R.~E.}\ \bibnamefont {Dietz}},\
  and\ \bibinfo {author} {\bibfnamefont {L.~R.}\ \bibnamefont {Walker}},\
  }\bibfield  {title} {\bibinfo {title} {Multiple magnon excitation in {NiO} by
  electron tunneling},\ }\href {https://doi.org/10.1103/physrevlett.27.1729}
  {\bibfield  {journal} {\bibinfo  {journal} {Physical Review Letters}\
  }\textbf {\bibinfo {volume} {27}},\ \bibinfo {pages} {1729} (\bibinfo {year}
  {1971})}\BibitemShut {NoStop}%
\bibitem [{\citenamefont {Moodera}\ \emph {et~al.}(1998)\citenamefont
  {Moodera}, \citenamefont {Nowak},\ and\ \citenamefont {van~de
  Veerdonk}}]{Moodera1998}%
  \BibitemOpen
  \bibfield  {author} {\bibinfo {author} {\bibfnamefont {J.~S.}\ \bibnamefont
  {Moodera}}, \bibinfo {author} {\bibfnamefont {J.}~\bibnamefont {Nowak}},\
  and\ \bibinfo {author} {\bibfnamefont {R.~J.~M.}\ \bibnamefont {van~de
  Veerdonk}},\ }\bibfield  {title} {\bibinfo {title} {Interface magnetism and
  spin wave scattering in ferromagnet-insulator-ferromagnet tunnel junctions},\
  }\href {https://doi.org/10.1103/physrevlett.80.2941} {\bibfield  {journal}
  {\bibinfo  {journal} {Physical Review Letters}\ }\textbf {\bibinfo {volume}
  {80}},\ \bibinfo {pages} {2941} (\bibinfo {year} {1998})}\BibitemShut
  {NoStop}%
\bibitem [{\citenamefont {Murai}\ \emph {et~al.}(1999)\citenamefont {Murai},
  \citenamefont {Ando}, \citenamefont {Kamijo}, \citenamefont {Kubota},\ and\
  \citenamefont {Miyazaki}}]{Murai1999}%
  \BibitemOpen
  \bibfield  {author} {\bibinfo {author} {\bibfnamefont {J.}~\bibnamefont
  {Murai}}, \bibinfo {author} {\bibfnamefont {Y.}~\bibnamefont {Ando}},
  \bibinfo {author} {\bibfnamefont {M.}~\bibnamefont {Kamijo}}, \bibinfo
  {author} {\bibfnamefont {H.}~\bibnamefont {Kubota}},\ and\ \bibinfo {author}
  {\bibfnamefont {T.}~\bibnamefont {Miyazaki}},\ }\bibfield  {title} {\bibinfo
  {title} {Direct observation of magnon excitation in a ferromagnetic tunnel
  junction using inelastic-electron-tunneling spectroscopy},\ }\href
  {https://doi.org/10.1143/jjap.38.l1106} {\bibfield  {journal} {\bibinfo
  {journal} {Japanese Journal of Applied Physics}\ }\textbf {\bibinfo {volume}
  {38}},\ \bibinfo {pages} {L1106} (\bibinfo {year} {1999})}\BibitemShut
  {NoStop}%
\bibitem [{\citenamefont {Fuchs}\ \emph {et~al.}(2004)\citenamefont {Fuchs},
  \citenamefont {Emley}, \citenamefont {Krivorotov}, \citenamefont {Braganca},
  \citenamefont {Ryan}, \citenamefont {Kiselev}, \citenamefont {Sankey},
  \citenamefont {Ralph}, \citenamefont {Buhrman},\ and\ \citenamefont
  {Katine}}]{Fuchs2004}%
  \BibitemOpen
  \bibfield  {author} {\bibinfo {author} {\bibfnamefont {G.~D.}\ \bibnamefont
  {Fuchs}}, \bibinfo {author} {\bibfnamefont {N.~C.}\ \bibnamefont {Emley}},
  \bibinfo {author} {\bibfnamefont {I.~N.}\ \bibnamefont {Krivorotov}},
  \bibinfo {author} {\bibfnamefont {P.~M.}\ \bibnamefont {Braganca}}, \bibinfo
  {author} {\bibfnamefont {E.~M.}\ \bibnamefont {Ryan}}, \bibinfo {author}
  {\bibfnamefont {S.~I.}\ \bibnamefont {Kiselev}}, \bibinfo {author}
  {\bibfnamefont {J.~C.}\ \bibnamefont {Sankey}}, \bibinfo {author}
  {\bibfnamefont {D.~C.}\ \bibnamefont {Ralph}}, \bibinfo {author}
  {\bibfnamefont {R.~A.}\ \bibnamefont {Buhrman}},\ and\ \bibinfo {author}
  {\bibfnamefont {J.~A.}\ \bibnamefont {Katine}},\ }\bibfield  {title}
  {\bibinfo {title} {Spin-transfer effects in nanoscale magnetic tunnel
  junctions},\ }\href {https://doi.org/10.1063/1.1781769} {\bibfield  {journal}
  {\bibinfo  {journal} {Applied Physics Letters}\ }\textbf {\bibinfo {volume}
  {85}},\ \bibinfo {pages} {1205} (\bibinfo {year} {2004})}\BibitemShut
  {NoStop}%
\bibitem [{\citenamefont {Zakeri}\ \emph
  {et~al.}(2023{\natexlab{c}})\citenamefont {Zakeri}, \citenamefont {Rau},
  \citenamefont {Wettstein}, \citenamefont {D\"ottling}, \citenamefont
  {Jandke}, \citenamefont {Yang}, \citenamefont {Wulfhekel},\ and\
  \citenamefont {Schmalian}}]{Zakeri2023a}%
  \BibitemOpen
  \bibfield  {author} {\bibinfo {author} {\bibfnamefont {K.}~\bibnamefont
  {Zakeri}}, \bibinfo {author} {\bibfnamefont {D.}~\bibnamefont {Rau}},
  \bibinfo {author} {\bibfnamefont {J.}~\bibnamefont {Wettstein}}, \bibinfo
  {author} {\bibfnamefont {M.}~\bibnamefont {D\"ottling}}, \bibinfo {author}
  {\bibfnamefont {J.}~\bibnamefont {Jandke}}, \bibinfo {author} {\bibfnamefont
  {F.}~\bibnamefont {Yang}}, \bibinfo {author} {\bibfnamefont {W.}~\bibnamefont
  {Wulfhekel}},\ and\ \bibinfo {author} {\bibfnamefont {J.}~\bibnamefont
  {Schmalian}},\ }\bibfield  {title} {\bibinfo {title} {Direct evidence of a
  charge depletion region at the interface of van der waals monolayers and
  dielectric oxides: The case of superconducting {FeSe}/{STO}},\ }\href
  {https://doi.org/10.1103/physrevb.107.184508} {\bibfield  {journal} {\bibinfo
   {journal} {Physical Review B}\ }\textbf {\bibinfo {volume} {107}},\ \bibinfo
  {pages} {184508} (\bibinfo {year} {2023}{\natexlab{c}})}\BibitemShut
  {NoStop}%
\bibitem [{\citenamefont {Balashov}\ \emph {et~al.}(2006)\citenamefont
  {Balashov}, \citenamefont {Tak{\'{a}}cs}, \citenamefont {Wulfhekel},\ and\
  \citenamefont {Kirschner}}]{Balashov2006}%
  \BibitemOpen
  \bibfield  {author} {\bibinfo {author} {\bibfnamefont {T.}~\bibnamefont
  {Balashov}}, \bibinfo {author} {\bibfnamefont {A.~F.}\ \bibnamefont
  {Tak{\'{a}}cs}}, \bibinfo {author} {\bibfnamefont {W.}~\bibnamefont
  {Wulfhekel}},\ and\ \bibinfo {author} {\bibfnamefont {J.}~\bibnamefont
  {Kirschner}},\ }\bibfield  {title} {\bibinfo {title} {Magnon excitation with
  spin-polarized scanning tunneling microscopy},\ }\href
  {https://doi.org/10.1103/physrevlett.97.187201} {\bibfield  {journal}
  {\bibinfo  {journal} {Physical Review Letters}\ }\textbf {\bibinfo {volume}
  {97}},\ \bibinfo {pages} {187201} (\bibinfo {year} {2006})}\BibitemShut
  {NoStop}%
\bibitem [{\citenamefont {Dzyapko}\ \emph {et~al.}(2013)\citenamefont
  {Dzyapko}, \citenamefont {Kurebayashi}, \citenamefont {Demidov},\ and\
  \citenamefont {Demokritov}}]{Dzyapko2013}%
  \BibitemOpen
  \bibfield  {author} {\bibinfo {author} {\bibfnamefont {O.}~\bibnamefont
  {Dzyapko}}, \bibinfo {author} {\bibfnamefont {H.}~\bibnamefont
  {Kurebayashi}}, \bibinfo {author} {\bibfnamefont {V.~E.}\ \bibnamefont
  {Demidov}},\ and\ \bibinfo {author} {\bibfnamefont {S.~O.}\ \bibnamefont
  {Demokritov}},\ }\bibfield  {title} {\bibinfo {title} {Control of pure spin
  current by magnon tunneling and three-magnon splitting in insulating yttrium
  iron garnet films},\ }in\ \href
  {https://doi.org/10.1016/b978-0-12-408130-7.00004-6} {\emph {\bibinfo
  {booktitle} {Recent Advances in Magnetic Insulators {\textendash} From
  Spintronics to Microwave Applications}}}\ (\bibinfo  {publisher} {Elsevier},\
  \bibinfo {year} {2013})\ pp.\ \bibinfo {pages} {83--122}\BibitemShut
  {NoStop}%
\bibitem [{\citenamefont {Ghazaryan}\ \emph {et~al.}(2018)\citenamefont
  {Ghazaryan}, \citenamefont {Greenaway}, \citenamefont {Wang}, \citenamefont
  {Guarochico-Moreira}, \citenamefont {Vera-Marun}, \citenamefont {Yin},
  \citenamefont {Liao}, \citenamefont {Morozov}, \citenamefont {Kristanovski},
  \citenamefont {Lichtenstein}, \citenamefont {Katsnelson}, \citenamefont
  {Withers}, \citenamefont {Mishchenko}, \citenamefont {Eaves}, \citenamefont
  {Geim}, \citenamefont {Novoselov},\ and\ \citenamefont
  {Misra}}]{Ghazaryan2018}%
  \BibitemOpen
  \bibfield  {author} {\bibinfo {author} {\bibfnamefont {D.}~\bibnamefont
  {Ghazaryan}}, \bibinfo {author} {\bibfnamefont {M.~T.}\ \bibnamefont
  {Greenaway}}, \bibinfo {author} {\bibfnamefont {Z.}~\bibnamefont {Wang}},
  \bibinfo {author} {\bibfnamefont {V.~H.}\ \bibnamefont {Guarochico-Moreira}},
  \bibinfo {author} {\bibfnamefont {I.~J.}\ \bibnamefont {Vera-Marun}},
  \bibinfo {author} {\bibfnamefont {J.}~\bibnamefont {Yin}}, \bibinfo {author}
  {\bibfnamefont {Y.}~\bibnamefont {Liao}}, \bibinfo {author} {\bibfnamefont
  {S.~V.}\ \bibnamefont {Morozov}}, \bibinfo {author} {\bibfnamefont
  {O.}~\bibnamefont {Kristanovski}}, \bibinfo {author} {\bibfnamefont {A.~I.}\
  \bibnamefont {Lichtenstein}}, \bibinfo {author} {\bibfnamefont {M.~I.}\
  \bibnamefont {Katsnelson}}, \bibinfo {author} {\bibfnamefont
  {F.}~\bibnamefont {Withers}}, \bibinfo {author} {\bibfnamefont
  {A.}~\bibnamefont {Mishchenko}}, \bibinfo {author} {\bibfnamefont
  {L.}~\bibnamefont {Eaves}}, \bibinfo {author} {\bibfnamefont {A.~K.}\
  \bibnamefont {Geim}}, \bibinfo {author} {\bibfnamefont {K.~S.}\ \bibnamefont
  {Novoselov}},\ and\ \bibinfo {author} {\bibfnamefont {A.}~\bibnamefont
  {Misra}},\ }\bibfield  {title} {\bibinfo {title} {Magnon-assisted tunnelling
  in van der waals heterostructures based on {CrBr}3},\ }\href
  {https://doi.org/10.1038/s41928-018-0087-z} {\bibfield  {journal} {\bibinfo
  {journal} {Nature Electronics}\ }\textbf {\bibinfo {volume} {1}},\ \bibinfo
  {pages} {344} (\bibinfo {year} {2018})}\BibitemShut {NoStop}%
\bibitem [{\citenamefont {Razdolski}\ \emph {et~al.}(2017)\citenamefont
  {Razdolski}, \citenamefont {Alekhin}, \citenamefont {Ilin}, \citenamefont
  {Meyburg}, \citenamefont {Roddatis}, \citenamefont {Diesing}, \citenamefont
  {Bovensiepen},\ and\ \citenamefont {Melnikov}}]{Razdolski2017}%
  \BibitemOpen
  \bibfield  {author} {\bibinfo {author} {\bibfnamefont {I.}~\bibnamefont
  {Razdolski}}, \bibinfo {author} {\bibfnamefont {A.}~\bibnamefont {Alekhin}},
  \bibinfo {author} {\bibfnamefont {N.}~\bibnamefont {Ilin}}, \bibinfo {author}
  {\bibfnamefont {J.~P.}\ \bibnamefont {Meyburg}}, \bibinfo {author}
  {\bibfnamefont {V.}~\bibnamefont {Roddatis}}, \bibinfo {author}
  {\bibfnamefont {D.}~\bibnamefont {Diesing}}, \bibinfo {author} {\bibfnamefont
  {U.}~\bibnamefont {Bovensiepen}},\ and\ \bibinfo {author} {\bibfnamefont
  {A.}~\bibnamefont {Melnikov}},\ }\bibfield  {title} {\bibinfo {title}
  {Nanoscale interface confinement of ultrafast spin transfer torque driving
  non-uniform spin dynamics},\ }\href@noop {} {\bibfield  {journal} {\bibinfo
  {journal} {Nature Communications}\ }\textbf {\bibinfo {volume} {8}},\
  \bibinfo {pages} {15007} (\bibinfo {year} {2017})}\BibitemShut {NoStop}%
\bibitem [{\citenamefont {Zakeri}\ and\ \citenamefont
  {Ernst}(2024)}]{Zakeri2024c}%
  \BibitemOpen
  \bibfield  {author} {\bibinfo {author} {\bibfnamefont {K.}~\bibnamefont
  {Zakeri}}\ and\ \bibinfo {author} {\bibfnamefont {A.}~\bibnamefont {Ernst}},\
  }\bibfield  {title} {\bibinfo {title} {Generation and propagation of
  ultrafast terahertz magnons in atomically architectured nanomagnets},\ }\href
  {https://doi.org/10.1021/acs.nanolett.4c01982} {\bibfield  {journal}
  {\bibinfo  {journal} {Nano Letters}\ }\textbf {\bibinfo {volume} {24}},\
  \bibinfo {pages} {9528} (\bibinfo {year} {2024})}\BibitemShut {NoStop}%
\end{thebibliography}%

\end{document}